\documentclass[12pt]{article}
\usepackage{epsfig}
\def\bra#1{\left\langle #1 \right|}
\def\ket#1{\left| #1 \right\rangle}
\def\vev#1{\left\langle #1 \right\rangle}
\begin{document}
\baselineskip=15.5pt
\renewcommand{\theequation}{\arabic{section}.\arabic{equation}}
\pagestyle{plain}
\setcounter{page}{1}
\def\appendix{
\par
\setcounter{section}{0}
\setcounter{subsection}{0}
\def\thesection{\Alph{section}}}
\begin{titlepage}

\leftline{\tt hep-th/0204012}

\vskip -.8cm

\rightline{\small{\tt CALT-68-2378}}
\rightline{\small{\tt CITUSC/02-011}}  

\begin{center}

\vskip 2 cm

{\LARGE {Open string states and D-brane tension}} 

\vskip .5cm

{\LARGE{from vacuum string field theory}}  

\vskip 1.5 cm

{\large Yuji Okawa}

\vskip 1 cm

{California Institute of Technology 452-48, \\
Pasadena, CA 91125, USA}

\smallskip 
{\tt okawa@theory.caltech.edu}

\vskip 2cm

{\bf Abstract}
\end{center}

\noindent
We propose a description of open string fields
on a D25-brane
in vacuum string field theory.
We show that the tachyon mass
is correctly reproduced from
our proposal and further argue that
the mass spectrum of all other open string states
is correctly obtained as well.
We identify the string coupling constant from
the three-tachyon coupling and show that
the tension of a D25-brane
is correctly expressed
in terms of the coupling constant,
which resolves the controversy in the literature.
We also discuss a reformulation
of our description which is rather similar to
boundary string field theory.

\end{titlepage}

\newpage


\section{Introduction}
\setcounter{equation}{0}

One of the most crucial open problems
regarding vacuum string field theory (VSFT)
\cite{Rastelli:2000hv}
is that while ratios of tensions of various D-branes
can be reproduced \cite{Rastelli:2001vb},
the tension of a single
D25-brane has not yet been obtained correctly.
Another important problem is that
we do not completely understand how to describe
open string states around D-branes
in the framework of VSFT.
The goal of this paper is to provide a resolution
to these two problems.

Actually, these two problems are closely related.
Since the D25-brane tension $T_{25}$ is related to
the on-shell three-tachyon coupling $g_T$ through
the relation \cite{Sen:1999xm, Ghoshal:2000gt}\footnote{
Since this relation plays an important role in this paper,
we verify it using the notation of \cite{Polchinski:rq}
in Appendix A. We use the convention that $\alpha'=1$
except in the introduction and Appendix A.
}
\begin{equation}
  T_{25} = \frac{1}{2 \pi^2 \alpha'^3 g_T^2},
\label{D25-brane-tension}
\end{equation}
the energy density ${\cal E}_c$ of the classical solution
corresponding to a single D25-brane
must satisfy
\begin{equation}
  \frac{{\cal E}_c}{T_{25}}
  = 2 \pi^2 \alpha'^3 g_T^2 ~ {\cal E}_c =1.
\label{E_c/T}
\end{equation}
However, the on-shell
three-tachyon coupling $g_T$
based on the earlier proposal
for the tachyon state \cite{Hata:2001sq}
failed to reproduce the relation (\ref{D25-brane-tension})
\cite{Hata:2001sq, Hata:2001wa},
and the ratio ${\cal E}_c/T_{25}$ turned out to be
\cite{Rastelli:2001wk, Hata:2002it}
\begin{equation}
  \frac{{\cal E}_c}{T_{25}}
  = \frac{\pi^2}{3} \left( \frac{16}{27 \ln 2} \right)^3
  \simeq 2.0558.
\label{wrong-E_c/T}
\end{equation}
This is regarded as
the most crucial problem with the earlier proposal
for the tachyon state \cite{Hata:2001sq}.
If we assume the universality of the ghost part of solutions
in VSFT \cite{Rastelli:2001jb},
the calculation of the ratio
${\cal E}_c/T_{25}$ involves only the matter sector of VSFT.
The matter part of the classical solution representing
a D25-brane is assumed to be described by the sliver state
\cite{Rastelli:2001jb}, which is one of the basic
assumptions in VSFT.
It is generally believed
that the wrong ratio (\ref{wrong-E_c/T})
is not due to the identification of the sliver state
as a D25-brane,
but originated in
the incorrect identification of the tachyon state.

Is there any canonical way to identify open string states
around a D-brane solution in VSFT?
Let us recall the situation in ordinary field theories.
Given a classical solution breaking translational invariance,
there must be a massless Goldstone mode around the solution.
This massless mode can be identified by infinitesimal
deformation of the collective coordinate.
If we apply the same logic to a lower-dimensional
D-brane solution in VSFT, the massless scalar field
on the D-brane should be identified
as the associated Goldstone mode.
Since a lower-dimensional D-brane is described
by a sliver state with a Dirichlet boundary condition
\cite{Rastelli:2001jb, Rastelli:2001vb, Mukhopadhyay:2001ey},
infinitesimal deformation of the
collective coordinate of the D-brane corresponds to
infinitesimal deformation of the Dirichlet
boundary condition.
As is familiar in the open string sigma model,
this is realized by
an insertion of the vertex operator of the massless scalar
integrated over the boundary of the sliver state.

This is easily generalized to other open string states.
In general, an open string state on a D-brane
is described by a sliver state
where the corresponding vertex operator
of the open string state
is integrated along the boundary
with the boundary condition
of the D-brane.
Actually, this idea has
already been discussed to some extent
in \cite{Rastelli:2001vb},
but little progress has been made thereafter.
We emphasize, however, that
it is crucial for this identification to work
for the consistency of a series of assumptions
regarding VSFT.
We will therefore revisit and explore this idea,
and show that the open string mass
spectrum and the D-brane tension are in fact
correctly reproduced.

Interestingly,
the resulting description of open string states
turns out to be rather close to
that of boundary string field theory (BSFT)
or background-independent open string field theory
\cite{Witten:1992qy, Witten:1992cr, Li:1993za,
Shatashvili:1993kk, Shatashvili:1993ps}.
In fact, our results are effectively reproduced
by a BSFT-like action.
We discuss some aspects of this reformulation
in this paper.

The organization of this paper is as follows.
In Section 2 we propose a description
of open string fields on a D25-brane in VSFT.
We then consider linearized equations of motion
for the open string states in Section 3
and explain how physical state conditions
such as the on-shell condition
are imposed.
In Section 4 we express the string field theory
action in terms of the open string fields
on a D25-brane.
We show that the kinetic terms of the open string fields
vanish when the fields satisfy the physical state conditions
despite some subtleties.
We then calculate
the normalization of the tachyon kinetic term
and the on-shell three-tachyon interaction
to identify the coupling constant $g_T$,
and show that it is correctly related to
the D25-brane tension $T_{25}$
through (\ref{D25-brane-tension}).
Section 5 is devoted to discussions
of some problematic issues in our formulation,
a BSFT-like reformulation
of our description,
and future directions.

\section{Open string states in vacuum string field theory}
\setcounter{equation}{0}

The action of VSFT is given by \cite{Rastelli:2000hv}
\begin{equation}
  S = -\frac{1}{2} \vev{\Psi | {\cal Q} | \Psi}
  -\frac{1}{3} \vev{\Psi | \Psi \ast \Psi},
\label{VSFT-action}
\end{equation}
where $\ket{\Psi}$ is the string field
represented by a state
with ghost number one
in the boundary conformal field theory (BCFT),
and where
the definitions of the BPZ inner product $\vev{A | B}$
and the star product $\ket{\Psi \ast \Psi}$
are standard ones \cite{Witten:1985cc}.
It is conjectured that
the state $\ket{\Psi}=0$ corresponds
to the tachyon vacuum,
and the operator ${\cal Q}$ with ghost number one
is made purely of ghost fields.
We defined $\ket{\Psi}$ and ${\cal Q}$
to absorb an overall normalization factor
including the open string coupling constant.
We take the matter part of the BCFT
to be the one describing a D25-brane.

Classical solutions corresponding to
various D-branes are assumed to take the factorized form,
\begin{equation}
  \ket{\Psi} = \ket{\Psi_g} \otimes \ket{\Psi_m},
\end{equation}
so that the equation of motion factorizes into
\begin{equation}
  {\cal Q} \ket{\Psi_g} + \ket{\Psi_g \ast \Psi_g} = 0,
\end{equation}
and
\begin{equation}
  \ket{\Psi_m} = \ket{\Psi_m \ast \Psi_m}.
\end{equation}
It is further assumed that
all D-brane solutions have the same ghost part.
On the other hand, the matter part is given by
the sliver state
with the boundary condition corresponding
to the D-brane \cite{Rastelli:2001vb}.

For example, a D25-brane is described by
the matter part of the sliver state $\ket{\Xi_m}$
with the Neumann boundary condition
defined by a limit of the matter part of
wedge states $\ket{n}$
\cite{Rastelli:2000iu},
\begin{equation}
  \ket{\Xi_m} = \lim_{n \to \infty} \ket{n}.
\label{sliver-limit}
\end{equation}
The matter part of the wedge states $\ket{n}$
is defined by
\begin{equation}
  \vev{ n | \phi} = {\cal N}
  \vev{f_n \circ \phi(0)}_{\rm UHP},
\label{sliver}
\end{equation}
for any state in the matter Fock space $\ket{\phi}$.
In (\ref{sliver}),
the conformal transformation $f_n (\xi)$ is given by
\begin{equation}
  f_n (\xi) = \frac{n}{2}
  \tan \left( \frac{2}{n} \tan^{-1} \xi \right),
\end{equation}
the correlation function is evaluated on the upper-half plane,
and ${\cal N}$ is an appropriate normalization factor.
Here the combination of the conformal transformation
$f_n (\xi)$ and the upper-half plane can be replaced
by any combination of a conformal transformation
$h_n (\xi)$ and Riemann surface $\Sigma_n$
as long as the combination is conformally equivalent
to that of $f_n (\xi)$ and the upper-half plane.

For lower-dimensional D$p$-branes, the Neumann sliver state
is replaced by the sliver state
with the Dirichlet boundary condition
for each of the $25-p$ transverse directions.
It is similarly defined as in (\ref{sliver})
by a limit of wedge states
with the following boundary condition for the string coordinate
$X^i (z)$ on the real axis $t$ of the upper-half plane:\footnote
{
For a more complete description, an appropriate regularization
is necessary. See \cite{Rastelli:2001vb} for details.
}
\begin{eqnarray}
  & \partial X^i (t) = \bar{\partial} X^i (t) & \quad
    {\rm for} \quad - \frac{n}{2} \tan \frac{\pi}{2 n} \le t
        \le \frac{n}{2} \tan \frac{\pi}{2 n},
\nonumber \\
  & X^i (t) = a^i & \quad
    {\rm for} \quad t < - \frac{n}{2} \tan \frac{\pi}{2 n},
              \quad t > \frac{n}{2} \tan \frac{\pi}{2 n},
\label{Dirichlet-sliver}
\end{eqnarray}
where $a^i$ is the position of the D-brane in space-time.
It is shown in \cite{Rastelli:2001vb}
that ratios of tensions of various D-branes
are correctly reproduced based on this description.

As we mentioned in the introduction, it is important
to identify open string states
around D-brane solutions correctly.
A proposal for the tachyon state around a D25-brane
in VSFT was put forward in the operator formulation 
\cite{Gross:1986ia, Gross:1986fk,
Cremmer:1986if, Samuel:1986qk, Ohta:wn}
by Hata and Kawano \cite{Hata:2001sq}.
Its conformal field theory (CFT) description
\cite{LeClair:1988sp, LeClair:1988sj} was
found to be the sliver state
with the tachyon vertex operator\footnote{
The normal ordering for vertex operators
is implicit throughout the paper.
}
$e^{ikX}$ inserted at the midpoint of the boundary
of the sliver state \cite{Rastelli:2001wk},
which corresponds to $t = \infty$ in our notation.
It turned out, however, that the D25-brane tension
is not correctly reproduced with this proposal
\cite{Hata:2001sq, Hata:2001wa, Rastelli:2001wk, Hata:2002it}.
Since it seems difficult to find the correct description
of the tachyon by trial and error, let us take
the approach suggested in the introduction.\footnote{
For other approaches
along the line of the CFT description \cite{Rastelli:2001wk}
of the Hata-Kawano tachyon state \cite{Hata:2001sq},
see \cite{Rashkov:2001js, Rashkov:2002xz, Matlock:2002dm}.
Possibilities of identifying D-branes
as classical solutions other than sliver-type configurations
were studied in \cite{Matsuo:2001yb, Kishimoto:2001de}.}

The massless scalar fields on a D-brane describing
its fluctuation in the transverse directions
are Goldstone modes associated with the broken
translational symmetries,
and should be identified
with infinitesimal deformations of the collective
coordinates.
Since the collective coordinates are encoded
as Dirichlet boundary conditions $X^i = a^i$
in (\ref{Dirichlet-sliver}), the massless scalar
fields must be identified with infinitesimal deformations
of the boundary condition.
It is well-known in the open string sigma model
that such a deformation
is realized by an insertion of an integral of
the vertex operator $\partial_\perp X^i e^{ikX}$
of the scalar field,
where $\partial_\perp$ is the derivative
normal to the boundary. Therefore, the massless scalar field
should be described by the sliver state
where the integral of the vertex operator is inserted
along the boundary
with the Dirichlet boundary condition.

This identification of the scalar fields is generalized to
other open string states using the relation between
the deformation of the boundary condition and the insertion
of an integrated vertex operator.
For example, the tachyon field $T(k)$ on a D25-brane
represented by the sliver state
$\ket{\Xi_m}$ (\ref{sliver-limit})
should be described at the linear order of $T(k)$
as follows:
\begin{equation}
  \ket{\Psi_m} = \ket{\Xi_m}
  - \int d^{26} k~ T(k) \ket{\chi_T (k)},
\end{equation}
where $\ket{\chi_T (k)}$ is defined for any state
in the matter Fock space $\ket{\phi}$ by
\begin{eqnarray}
  \vev{\chi_T (k) | \phi}
  = \lim_{n \to \infty} {\cal N}
  \vev{\int dt~ e^{ikX(t)}
  h_n \circ \phi(0)}_{\Sigma_n}.
\label{chi_T}
\end{eqnarray}
Here a combination of a conformal transformation
$h_n (\xi)$ and Riemann surface $\Sigma_n$
should be conformally equivalent to that of
$f_n (\xi)$ and the upper-half plane,
and the integral of the tachyon vertex operator $e^{ikX(t)}$
is taken along the boundary of the wedge state from
$h_n (1)$ to $h_n (-1)$.\footnote{
In the case of the upper-half plane, for example,
the integral is taken along
$$
  \int_{f_n (1)}^{\infty} dt
  + \int_{-\infty}^{f_n (-1)} dt
  = \int_{\frac{n}{2} \tan \frac{\pi}{2 n}}^{\infty} dt
  + \int_{-\infty}^{-\frac{n}{2} \tan \frac{\pi}{2 n}} dt.
$$
}
Since the integrated vertex operator is not generically
conformally invariant, the definition of the tachyon state
depends on the choice of ($h_n (\xi)$, $\Sigma_n$).
Note, however, that this ambiguity is absent if
the tachyon is on shell, $k^2=1$, because the integral of
the vertex operator becomes conformally invariant.
Therefore, the ambiguity coming from the choice of
($h_n (\xi)$, $\Sigma_n$) can be regarded
as that of field redefinition of the tachyon field.

This ambiguity can also be described as follows.
We can make a conformal transformation such that
the Riemann surface $\Sigma_n$,
where the off-shell tachyon field is defined
in (\ref{chi_T}),
is mapped to a cone subtending an angle $n \pi$
at the origin. Its boundary is parametrized
as $e^{i \theta}$
with $-n \pi/2 \le \theta \le n \pi/2$,
and we choose the region of the integral
of the vertex operator to be
$-(n-1) \pi/2 \le \theta \le (n-1) \pi/2$.
\footnote{
In other words, the boundary
of the local coordinate $\xi$ \cite{Rastelli:2001vb},
which is $-1 \le \xi \le 1$ on the real axis, is mapped
to the sum of the two regions
$(n-1) \pi/2 \le \theta \le n \pi/2$
and $-n \pi/2 \le \theta \le -(n-1) \pi/2$.
}
The inserted operator now takes the form
\begin{equation}
  \int_{-(n-1) \pi/2}^{(n-1) \pi/2} d \theta~
  {\cal F}_0 (\theta)^{k^2-1} e^{ikX (e^{i \theta})},
\label{cone-integral}
\end{equation}
where the additional factor ${\cal F}_0 (\theta)^{k^2-1}$
comes from the conformal transformation from $\Sigma_n$
to the cone. The ambiguity of the off-shell definition
of the tachyon is now encoded in this factor.
We will frequently use this representation of off-shell
tachyon configurations in the rest of the paper and refer
to this as {\it the cone representation}.

So far we have considered infinitesimal deformations
of the BCFT.
It is easily generalized to finite deformations
as follows:
\begin{equation}
  \vev{ \{ \varphi_i \} | \phi}
  = \lim_{n \to \infty} {\cal N}
  \vev{\exp \left[ - \int dt \int d^{26} k
  \sum_i \varphi_i (k)
  {\cal O}_{\varphi_i (k)} (t) \right]
  h_n \circ \phi(0)}_{\Sigma_n},
\label{general}
\end{equation}
where $\{ \varphi_i \}$ denotes the open string fields
on a D25-brane such as the tachyon $T(k)$ or the massless
gauge field $A_\mu (k)$ collectively,
${\cal O}_{\varphi_i (k)}$ is the vertex operator corresponding
to the field $\varphi_i (k)$, and the integral over $t$
is taken along the boundary as before.
This is a formal definition because
we need to regularize divergences which appear
when some of the operators ${\cal O}_{\varphi_i (k)}$
coincide. We will come back to this point later
in Sections 4 and 5.
There would also be some ambiguity
when we assign vertex operators
${\cal O}_{\varphi_i (k)}$ to fields $\varphi_i (k)$
when they are off shell.
It would be natural, however,
to associate $e^{ikX}$ with the tachyon
because it is primary
even when the momentum $k$ is off shell.

\section{Equations of motion for open string states}
\setcounter{equation}{0}

Let us consider the linearized
equations of motion
for the open string fields $\{ \varphi_i \}$
to see if our identification of the open string states
on a single D25-brane works:

\begin{equation}
  \ket{\chi_{\varphi_i} (k)}
  = \ket{\chi_{\varphi_i} (k) \ast \Xi_m}
  + \ket{\Xi_m \ast \chi_{\varphi_i} (k)},
\label{equation-of-motion}
\end{equation}
where
\begin{eqnarray}
  \vev{\chi_{\varphi_i} (k) | \phi}
  = \lim_{n \to \infty} {\cal N}
  \vev{\int dt~ {\cal O}_{\varphi_i} (t)~
  h_n \circ \phi(0)}_{\Sigma_n},
\end{eqnarray}
for any state in the matter Fock space $\ket{\phi}$
and the integral over $t$ is the same as (\ref{chi_T}).

As we will see in more detail shortly,
the sum of the two integrals of the vertex operator
on the right-hand side of (\ref{equation-of-motion})
gives the integral on the left-hand side
in the large $n$ limit.
Therefore, it is easily understood
that the equation of motion
(\ref{equation-of-motion}) can be satisfied.
What is less obvious is how the physical state conditions
are imposed. For example, the on-shell condition $k^2=1$
must be imposed for the tachyon,
and the massless condition $k^2=0$ and
the transversality condition must be imposed
for the gauge field.
Let us take a closer look at the large $n$ limit
by considering
the following inner products ${\cal A}_L$ and ${\cal A}_R$
defined by
\begin{equation}
  {\cal A}_L = \vev{ \phi | \chi_{\varphi_i} (k)}_n, \qquad
  {\cal A}_R = \bra{\phi}
  ( \ket{\chi_{\varphi_i} (k)}_n \ast \ket{n}
  + \ket{n} \ast \ket{\chi_{\varphi_i} (k)}_n ),
\end{equation}
where $\ket{\phi}$ is an arbitrary state
in the matter Fock space
and the large $n$ limit is not taken for the open string state
$\ket{\chi_{\varphi_i} (k)}_n$
as is indicated by the subscript $n$.
The equation of motion (\ref{equation-of-motion})
contracted with $\bra{\phi}$ is given by
\begin{equation}
  \lim_{n \to \infty} {\cal A}_L
  = \lim_{n \to \infty} {\cal A}_R.
\end{equation}

Let us first consider the case
without the integral of the vertex operator
along the boundary.
The inner products ${\cal A}_L$ and ${\cal A}_R$
then reduce to
\begin{equation}
  {\cal A}_L \to \vev{ \phi | n }, \quad
  {\cal A}_R \to 2 \vev{ \phi | n \ast n }
  = 2 \vev{ \phi | 2n-1 },
\end{equation}
where we used the famous star algebra of wedge states
\cite{Rastelli:2000iu}:
\begin{equation}
  \ket{n} \ast \ket{m} = \ket{n+m-1}.
\end{equation}
What is important here is
that the two conformal transformations
$f_n (\xi)$ and $f_{2n-1} (\xi)$ used in defining $\ket{n}$
and $\ket{2n-1}$, respectively, have the same large $n$ limit:
\begin{equation}
  \lim_{n \to \infty} f_n (\xi) 
  = \lim_{n \to \infty} f_{2n-1} (\xi) = \tan^{-1} \xi.
\end{equation}
This was essential for the sliver state
to solve the matter equation of motion of VSFT
\cite{Rastelli:2001vb}.

Now consider the effect of the integrated vertex operator.
Let us take the case of the tachyon
as an example.
The inner product ${\cal A}_L$ is given by
\begin{eqnarray}
  {\cal A}_L = \vev{\phi | \chi_T (k)}_n
  = {\cal N}
  \vev{\int_{C} dt~ e^{ikX (t)} {\cal F}_L (t)^{k^2-1}~
  f_n \circ \phi(0)}_{\rm UHP},
\end{eqnarray}
where the contour $C$ is given by
\begin{equation}
  \int_C dt
  = \int_{f_n (1)}^{\infty} dt
  + \int_{-\infty}^{f_n (-1)} dt
  = \int_{\frac{n}{2} \tan \frac{\pi}{2 n}}^{\infty} dt
  + \int_{-\infty}^{-\frac{n}{2} \tan \frac{\pi}{2 n}} dt.
\label{contour}
\end{equation}
Note that we inserted a factor ${\cal F}_L (t)^{k^2-1}$
which comes from a conformal transformation from
the Riemann surface $\Sigma_n$, where the off-shell tachyon
is defined, to the upper-half plane.
In constructing the states $\ket{\chi_T (k)}_n \ast \ket{n}$
and $\ket{n} \ast \ket{\chi_T (k)}_n$, the vertex operator
undergoes a further conformal transformation.
The contour $C$ (\ref{contour}) is mapped to
\begin{equation}
  \int_{C_R} dt
  = \int_{f_{2n-1} (1)}^{\infty} dt
  = \int_{\frac{2n-1}{2} \tan \frac{\pi}{2 (2n-1)}}^{\infty} dt
\label{right-contour}
\end{equation}
for $\ket{\chi_T (k)}_n \ast \ket{n}$, and to
\begin{equation}
  \int_{C_L} dt
  = \int_{-\infty}^{f_{2n-1} (-1)} dt
  = \int_{-\infty}^{-\frac{2n-1}{2} \tan \frac{\pi}{2 (2n-1)}} dt
\label{left-contour}
\end{equation}
for $\ket{n} \ast \ket{\chi_T (k)}_n$.
The inner product ${\cal A}_R$ is then given by
\begin{eqnarray}
  {\cal A}_R &=& \bra{\phi}
  ( \ket{\chi_T (k)}_n \ast \ket{n}
  + \ket{n} \ast \ket{\chi_T (k)}_n )
\nonumber \\
  &=& {\cal N}
  \vev{\int_{C'} dt~ e^{ikX (t)} {\cal F}_R (t)^{k^2-1}~
  f_{2n-1} \circ \phi(0)}_{\rm UHP},
\end{eqnarray}
where the contour $C'$
is the sum of $C_R$ (\ref{right-contour}) and
$C_L$ (\ref{left-contour}).
Note that the factor ${\cal F}_R (t)^{k^2-1}$ in ${\cal A}_R$
is different from the one
${\cal F}_L (t)^{k^2-1}$ in ${\cal A}_L$
because of the additional conformal transformations.
Therefore,
although $f_n (\xi)$ and $f_{2n-1} (\xi)$ have the same
large $n$ limit and the contours $C$ and $C'$
become the same in the large $n$ limit,
the large $n$ limit of ${\cal A}_L$ and
that of ${\cal A}_R$ do not coincide
because of the difference between
${\cal F}_L (t)^{k^2-1}$ and ${\cal F}_R (t)^{k^2-1}$
{\it unless} the condition $k^2=1$ is satisfied.
This shows that the tachyon state $\ket{\chi_T (k)}$
satisfies the equation of motion only when $k^2=1$,
which is the correct on-shell condition for the tachyon. 

The origin of the condition $k^2=1$ is obvious: it is
the condition that the integral of the vertex operator
$e^{ikX}$ is conformally invariant.
In other words, the vertex operator must be
a primary field with conformal dimension one,
which is nothing but the physical state condition
for the vertex operator in string theory.
The same argument applies to all other open string states:
the equation of motion is satisfied when the vertex operator
${\cal O}_{\varphi_i (k)}$ is a primary field
with conformal dimension one.
For the gauge field,
this imposes the transversality condition
as well as the on-shell condition.
The vertex operator for the gauge
field $\zeta_\mu \partial_t X^\mu e^{ikX}$ does not
transform as a tensor unless the transversality
condition $\zeta \cdot k=0$ is satisfied.
The non-tensor property would result in different
expressions for ${\cal A}_L$ and ${\cal A}_R$
which violate the equation of motion
as in the case of an off-shell momentum.

\section{String field theory action
in terms of open string fields}
\setcounter{equation}{0}

Based on our proposal for the description
of the open string fields
$\ket{\{ \varphi_i \}}$ (\ref{general}),
we can rewrite the VSFT action
in terms of the open string fields
$\{ \varphi_i \}$ as follows:
\begin{equation}
  S \left[ \{ \varphi_i \} \right]
  = - \vev{\Psi_g | {\cal Q} | \Psi_g}
  \left[ \frac{1}{2} \vev{ \{ \varphi_i \} | \{ \varphi_i \} }
  - \frac{1}{3} \vev{ \{ \varphi_i \} |
  \{ \varphi_i \} \ast \{ \varphi_i \} } \right].
\label{varphi-action}
\end{equation}

The kinetic terms of the open string fields
might be expected to be reproduced correctly
in view of the argument presented in the previous section.
However, it is not automatic.
We have shown that the states $\ket{\chi_{\varphi_i} (k)}$
satisfy the equations of motion (\ref{equation-of-motion})
when contracted
with any state in the matter Fock space $\ket{\phi}$.
The proof can be generalized to cases where
the equations of motion (\ref{equation-of-motion}) are
contracted with a larger class of states such as
wedge states with some operators inserted
as long as the size of the wedge stays finite
while we take the large $n$ limit for
$\ket{\chi_{\varphi_i} (k)}$ and $\ket{\Xi_m}$.
When we evaluate the kinetic terms of the fields $\{ \varphi_i \}$
in (\ref{varphi-action}), however,
we have to handle the following combination
of inner products:
\begin{equation}
  \vev{ \chi_{\varphi_i} (-k) | \chi_{\varphi_i} (k) }
  - \vev{ \chi_{\varphi_i} (-k) | \chi_{\varphi_i} (k) \ast \Xi_m }
  - \vev{ \chi_{\varphi_i} (-k) | \Xi_m \ast \chi_{\varphi_i} (k) }.
\end{equation}
This takes the form of the equation of motion
(\ref{equation-of-motion}) contracted
with $\bra{\chi_{\varphi_i} (-k)}$, but we cannot apply
the argument in the previous section to this case
because of two subtleties.
First, the expression contains divergences when
two vertex operators coincide.
Secondly, we have to take the large $n$ limit for
$\bra{\chi_{\varphi_i} (-k)}$ as well so that 
the state $\bra{\chi_{\varphi_i} (-k)}$ does not belong to
the class of states we just mentioned.
Therefore, it is important to verify
whether the correct kinetic
terms are reproduced for the consistency of our proposal.
In fact, it was argued in \cite{Rastelli:2001wk}
that this is where the earlier proposal \cite{Hata:2001sq}
for the tachyon state failed.
In Subsection 4.1
we will show that
the kinetic terms of the fields $\{ \varphi_i \}$
in (\ref{varphi-action}) vanish
when they satisfy the physical state conditions.

As we mentioned in the introduction, we are particularly
interested in the on-shell three-tachyon coupling $g_T$
which is related to the D25-brane tension through
(\ref{D25-brane-tension}).
We will calculate
the normalization of the tachyon kinetic term
in Subsection 4.2
and the on-shell three-tachyon interaction
in Subsection 4.3
to show that the correct D25-brane tension is actually
reproduced from (\ref{varphi-action}).

\subsection{Open string mass spectrum}

Let us begin with the tachyon
and set other fields to zero
in $\ket{\{ \varphi_i \}}$ (\ref{general}) for simplicity.
If we denote the resulting state
representing tachyon field configurations
by $| e^{-T} \rangle$,
\begin{equation}
  \langle e^{-T} | \phi \rangle
  = \lim_{n \to \infty} {\cal N}
  \vev{\exp \left[ - \int dt \int d^{26} k~ T(k)
  e^{ikX} (t) \right]
  h_n \circ \phi(0)}_{\Sigma_n},
\label{finite-tachyon}
\end{equation}
the action for the tachyon field is given by
\begin{equation}
  S \left[ T(k) \right]
  = - \vev{\Psi_g | {\cal Q} | \Psi_g}
  \left[ \frac{1}{2} \langle e^{-T} | e^{-T} \rangle
  - \frac{1}{3} \langle e^{-T} |
  e^{-T} \ast e^{-T} \rangle \right].
\label{tachyon-action}
\end{equation}
The state $| e^{-T} \rangle$ has nonlinear dependence on
the tachyon field $T(k)$. If we expand it in powers of
$T(k)$, we have
\begin{equation}
  | e^{-T} \rangle
  = \ket{\it 0} + \ket{\it 1} + \ket{\it 2}
  + \ket{\it 3} + \ldots,
\end{equation}
where\footnote{
These states with italic numbers,
$\ket{\it 0}, \ket{\it 1}, \ket{\it 2}, \ldots$,
should not be confused with wedge states
$\ket{0}, \ket{1}, \ket{2}, \ldots$.
}
\begin{eqnarray}
  \ket{\it 0} = \ket{\Xi_m}, \quad
  \ket{\it 1} = - \int d^{26} k~ T(k) \ket{\chi_T (k)},
  \ldots.
\end{eqnarray}
The tachyon kinetic term is given by
\begin{equation}
  S^{(2)} = - \vev{\Psi_g | {\cal Q} | \Psi_g}
  \left[ \frac{1}{2} \vev{ {\it 1} | {\it 1} }
         + \vev{ {\it 2} | {\it 0} }
         - \vev{ {\it 1} | {\it 1} \ast {\it 0} }
         - \vev{ {\it 2} | {\it 0} \ast {\it 0} }
  \right].
\label{S^2}
\end{equation}
This takes the following form:
\begin{equation}
  S^{(2)} = -\frac{\cal K}{2} (2 \pi)^{26}
  \int d^{26} k~ K(k^2) T(k) T(-k),
\end{equation}
where we denote the density of
$\vev{\Xi_m | \Xi_m}
\vev{\Psi_g | {\cal Q} | \Psi_g}$ by ${\cal K}$,
namely,
\begin{equation}
  \vev{\Xi_m | \Xi_m} \vev{\Psi_g | {\cal Q} | \Psi_g}
  = \int d^{26} x~{\cal K},
\end{equation}
and $K(k^2)$ consists of contributions from
$\vev{ {\it 1} | {\it 1} }$, $\vev{ {\it 2} | {\it 0} }$,
$\vev{ {\it 1} | {\it 1} \ast {\it 0} }$,
and $\vev{ {\it 2} | {\it 0} \ast {\it 0} }$,
\begin{equation}
  \frac{1}{2} K(k^2) = \frac{1}{2} K_{11}(k^2) + K_{20}(k^2)
                       - K_{110}(k^2) - K_{200}(k^2).
\label{K-from-4-terms}
\end{equation}
It is naively expected that
$\vev{ {\it 2} | {\it 0} }
- \vev{ {\it 2} | {\it 0} \ast {\it 0} }$
vanishes because of the equation of motion
for the sliver state,
$\ket{ {\it 0} \ast {\it 0} } = \ket{\it 0}$,
and the remaining combination of terms,
\begin{equation}
  \frac{1}{2} \vev{ {\it 1} | {\it 1} }
  - \vev{ {\it 1} | {\it 1} \ast {\it 0} }
  = \frac{1}{2} \left[ \vev{ {\it 1} | {\it 1} }
  - \vev{ {\it 1} | {\it 1} \ast {\it 0} }
  - \vev{ {\it 1} | {\it 0} \ast {\it 1} }
  \right],
\end{equation}
vanishes when the tachyon is on shell because of
the equation of motion for $\ket{\chi_T (k)}$.
However, it turns out that the story is more complicated
because of the subtleties we mentioned at the beginning of
this section.

Let us begin with the calculations
of $\vev{ {\it 1} | {\it 1} }$
and $\vev{ {\it 1} | {\it 1} \ast {\it 0} }$.
To avoid the singularity which arises when the two
vertex operators coincide, we regularize
the state $\ket{\it 1}$. We use the cone representation
of $\ket{\it 1}$ explained in the paragraph containing
(\ref{cone-integral}) and regularize the integral
of the vertex operator as follows:
\begin{equation}
  \int_{-(n-1) \pi/2 +\epsilon_0/2}^{(n-1) \pi/2
  -\epsilon_0/2} d \theta~
  {\cal F}_0 (\theta)^{k^2-1} e^{ikX (e^{i \theta})}.
\end{equation}
The inner product $\vev{\phi | {\it 1}}$ is expressed
on a cone with an angle $n \pi$.
To construct the inner product $\vev{{\it 1} | {\it 1}}$,
we cut off the region of the cone where the local coordinate
is mapped, which leaves a sector of an angle $(n-1) \pi$.
By gluing two such sectors together, $\vev{{\it 1} | {\it 1}}$
is expressed on a cone with an angle $2(n-1) \pi$.
We can map it to the unit disk by the conformal transformation
$z^{1/(n-1)}$.
Using the propagator on the unit disk,
\begin{equation}
  \vev{ e^{ikX(e^{i \theta_1})}
        e^{ik'X(e^{i \theta_2})} }_{\rm disk}
  = (2 \pi)^{26} \delta (k+k')
    \left| 2 \sin \frac{\theta_2 - \theta_1}{2}
    \right|^{2 k k'},
\label{disk-propagator}
\end{equation}
$K_{11}(k^2)$ is given by
\begin{eqnarray}
  && K_{11}(k^2)
  = \int_{-\pi/2+\epsilon/2}^{\pi/2-\epsilon/2} d \theta_2
  \int_{-\pi}^{-\pi/2-\epsilon/2} d \theta_1~
  {\cal F}(\theta_1+\pi)^{k^2-1} {\cal F} (\theta_2)^{k^2-1}
  \left| 2 \sin \frac{\theta_2 - \theta_1}{2}
  \right|^{-2 k^2}
\nonumber \\
  && \qquad
  + \int_{-\pi/2+\epsilon/2}^{\pi/2-\epsilon/2} d \theta_2
  \int_{\pi/2+\epsilon/2}^{\pi} d \theta_1~
  {\cal F}(\theta_1-\pi)^{k^2-1} {\cal F} (\theta_2)^{k^2-1}
  \left| 2 \sin \frac{\theta_2 - \theta_1}{2}
  \right|^{-2 k^2},
\label{K_11}
\end{eqnarray}
where
\begin{equation}
  \epsilon = \frac{\epsilon_0}{n-1}, \qquad
  {\cal F} (\theta)
  = \frac{1}{n-1} {\cal F}_0 ( (n-1) \theta ).
\label{F_0-to-F}
\end{equation}
The difference between ${\cal F}_0$ and ${\cal F}$
comes from the conformal transformation $z^{1/(n-1)}$.
Since $\epsilon$ goes to zero
when we take the large $n$ limit,
we do not need to take the limit $\epsilon_0 \to 0$
as in the case of the similar regularizations
discussed in \cite{Rastelli:2001vb}.

The construction of $\vev{ {\it 1} | {\it 1} \ast {\it 0} }$
can be done in a similar way.
By gluing together
three sectors with an angle $(n-1) \pi$ coming from
two $\ket{\it 1}$'s and one $\ket{\it 0}$,
$\vev{ {\it 1} | {\it 1} \ast {\it 0} }$ is expressed
on a cone with an angle $3(n-1) \pi$.
We can make a conformal transformation so that the cone
is mapped to the unit disk. However, it is more convenient
to make the same conformal transformation $z^{1/(n-1)}$
as the case of $\vev{{\it 1} | {\it 1}}$, which maps
the cone with an angle $3(n-1) \pi$ to a cone with
an angle $3 \pi$.
The propagator on this cone is given by
\begin{equation}
  \vev{ e^{ikX(e^{i \theta_1})}
        e^{ik'X(e^{i \theta_2})} }_{3 \pi}
  = (2 \pi)^{26} \delta (k+k')
    \left| 3 \sin \frac{\theta_2 - \theta_1}{3}
    \right|^{2 k k'},
\label{cone-propagator}
\end{equation}
which respects the periodicity
$\theta_i = \theta_i + 3 \pi$,
and $K_{110}(k^2)$ is given by
\begin{equation}
  K_{110}(k^2)
  = \int_{-\pi/2+\epsilon/2}^{\pi/2-\epsilon/2} d \theta_2
  \int_{-3 \pi/2+\epsilon/2}^{-\pi/2-\epsilon/2} d \theta_1~
  {\cal F}(\theta_1+\pi)^{k^2-1} {\cal F} (\theta_2)^{k^2-1}
  \left| 3 \sin \frac{\theta_2 - \theta_1}{3}
  \right|^{-2 k^2}.
\label{K_110}
\end{equation}

Let us calculate $K_{11}(k^2)/2-K_{110}(k^2)$
when the tachyon is on shell, $k^2=1$,
to see if it actually vanishes as naively expected.
The expressions for $K_{11}(k^2)$ and $K_{110}(k^2)$
simplify when $k^2=1$ so that
the integrals in $K_{11}(1)$ and $K_{110}(1)$
are easily performed to give
\begin{eqnarray}
  && \frac{1}{2} K_{11}(1)
  = \frac{1}{2}
  \int_{-\pi/2+\epsilon/2}^{\pi/2-\epsilon/2} d \theta_2
  \int_{-\pi}^{-\pi/2-\epsilon/2} d \theta_1~
  \left| 2 \sin \frac{\theta_2 - \theta_1}{2}
  \right|^{-2}
\nonumber \\
  && \qquad  \qquad \quad
  + \frac{1}{2}
  \int_{-\pi/2+\epsilon/2}^{\pi/2-\epsilon/2} d \theta_2
  \int_{\pi/2+\epsilon/2}^{\pi} d \theta_1~
  \left| 2 \sin \frac{\theta_2 - \theta_1}{2}
  \right|^{-2}
\nonumber \\
  && \qquad \qquad
  = - \ln \sin \frac{\epsilon}{2}
  = - \ln \epsilon + \ln 2 + O(\epsilon^2),
\end{eqnarray}
and
\begin{eqnarray}
  && K_{110}(1)
  = \int_{-\pi/2+\epsilon/2}^{\pi/2-\epsilon/2} d \theta_2
  \int_{-3 \pi/2+\epsilon/2}^{-\pi/2-\epsilon/2} d \theta_1~
  \left| 3 \sin \frac{\theta_2 - \theta_1}{3}
  \right|^{-2}
\nonumber \\
  && \qquad \qquad
  = - \ln \sin \frac{\epsilon}{3} + 2 \ln \sin \frac{\pi}{3}
  - \ln \sin \left( \frac{\pi}{3} + \frac{\epsilon}{3} \right)
\nonumber \\
  && \qquad \qquad
  = - \ln \epsilon + \ln \frac{3 \sqrt{3}}{2} + O(\epsilon).
\end{eqnarray}
Therefore, the divergent part in
$K_{11}(1)/2-K_{110}(1)$ vanishes,
but the finite part remains:
\begin{equation}
  \frac{1}{2} K_{11}(1) - K_{110}(1)
  = - \ln \frac{3 \sqrt{3}}{4} + O(\epsilon).
\label{finite-term-1}
\end{equation}
Does this imply the breakdown
of the tachyon equation of motion?

Recall, however, that there are other contributions
to the tachyon kinetic term, namely, $K_{20}(k^2)$
and $K_{200}(k^2)$ in (\ref{K-from-4-terms}).
Let us calculate them.
The state $\ket{\it 2}$ needs to be regularized.
We regularize the integrals of the inserted vertex operators
in the cone representation as follows:
\begin{equation}
  \int_{-(n-1) \pi/2 +3 \epsilon_0/2}^{(n-1) \pi/2
  -\epsilon_0/2} d \theta_2
  \int_{-(n-1) \pi/2 +\epsilon_0/2}^{\theta_2-\epsilon_0}
  d \theta_1~
  {\cal F}_0 (\theta_1)^{k^2-1} e^{ikX (e^{i \theta_1})}
  {\cal F}_0 (\theta_2)^{k^2-1} e^{ikX (e^{i \theta_2})}.
\end{equation}
The inner products $\vev{ {\it 2} | {\it 0} }$
and $\vev{ {\it 2} | {\it 0} \ast {\it 0} }$
are constructed similarly as in the cases
of $\vev{ {\it 1} | {\it 1} }$
and $\vev{ {\it 1} | {\it 1} \ast {\it 0} }$.
The resulting expressions for
$K_{20}(k^2)$ and $K_{200}(k^2)$
are given by
\begin{equation}
  K_{20}(k^2)
  = \int_{-\pi/2+3\epsilon/2}^{\pi/2-\epsilon/2} d \theta_2
  \int_{-\pi/2+\epsilon/2}^{\theta_2-\epsilon} d \theta_1~
  {\cal F}(\theta_1)^{k^2-1} {\cal F} (\theta_2)^{k^2-1}
  \left| 2 \sin \frac{\theta_2 - \theta_1}{2}
  \right|^{-2 k^2},
\label{K_20}
\end{equation}
and
\begin{equation}
  K_{200}(k^2)
  = \int_{-\pi/2+3\epsilon/2}^{\pi/2-\epsilon/2} d \theta_2
  \int_{-\pi/2+\epsilon/2}^{\theta_2-\epsilon} d \theta_1~
  {\cal F}(\theta_1)^{k^2-1} {\cal F} (\theta_2)^{k^2-1}
  \left| 3 \sin \frac{\theta_2 - \theta_1}{3}
  \right|^{-2 k^2}.
\label{K_200}
\end{equation}
Their on-shell values, $K_{20}(1)$ and $K_{200}(1)$,
are again easily calculated. Since
\begin{eqnarray}
  && \int_{-\pi/2+3\epsilon/2}^{\pi/2-\epsilon/2} d \theta_2
  \int_{-\pi/2+\epsilon/2}^{\theta_2-\epsilon} d \theta_1~
  \left| 2 n \sin \frac{\theta_2 - \theta_1}{2 n}
  \right|^{-2}
\nonumber \\
  && \qquad
  = \frac{\pi - 2 \epsilon}{2 n} \cot \frac{\epsilon}{2 n}
  + \ln \sin \frac{\epsilon}{2 n}
  - \ln \sin \left( \frac{\pi}{2 n} - \frac{\epsilon}{2 n} \right)
\nonumber \\
  && \qquad
  = \frac{\pi}{\epsilon} + \ln \epsilon
  - \ln \left( 2 n \sin \frac{\pi}{2 n} \right) -2
  + O(\epsilon),
\end{eqnarray}
for $n \ge 1 $, we have
\begin{eqnarray}
  K_{20}(1)
  &=& \frac{\pi}{\epsilon} + \ln \epsilon
  - \ln 2 -2
  + O(\epsilon),
\label{on-shell-K_20}
\\
  K_{200}(1)
  &=& \frac{\pi}{\epsilon} + \ln \epsilon
  - \ln \frac{3 \sqrt{3}}{2} -2
  + O(\epsilon).
\end{eqnarray}
The divergent part in $K_{20}(1)-K_{200}(1)$ vanishes,
but the finite part again remains:
\begin{equation}
  K_{20}(1) - K_{200}(1) = \ln \frac{3 \sqrt{3}}{4} + O(\epsilon).
\label{finite-term-2}
\end{equation}
However, this precisely cancels $K_{11}(1)/2-K_{110}(1)$
in the limit $\epsilon \to 0$.
Therefore, the tachyon kinetic term vanishes when $k^2=1$
in the large $n$ limit,
\begin{equation}
  \frac{1}{2} K(1) = \frac{1}{2} K_{11}(1) + K_{20}(1)
  - K_{110}(1) - K_{200}(1) = O(\epsilon),
\label{vanishing-kinetic-term}
\end{equation}
which is the property we expect for the correct
tachyon kinetic term.

The cancellation of the finite terms
in $K_{11}(1)/2 - K_{110}(1)$ and $K_{20}(1) - K_{200}(1)$
may seem rather accidental
and, apparently,
it seems to have nothing to do with
the argument for the mass spectrum
in Section 3 based on the conformal
invariance of the integrated vertex operator.
However, 
the cancellation can be regarded as a consequence
of the conformal property of the vertex operators,
as we will show in Appendix C
using some results from
Appendix B.
Note also that the finite term in (\ref{finite-term-2})
does not depend on details of the regularization
of the state $\ket{\it 2}$.
We obtain the same result if we regularize the integrals
in (\ref{K_20}) and (\ref{K_200}) as
\begin{equation}
  \int_{-\pi/2+\epsilon+\eta}^{\pi/2-\eta} d \theta_2
  \int_{-\pi/2+\eta}^{\theta_2-\epsilon} d \theta_1
\end{equation}
as long as $\epsilon$ and $\eta$ go to zero in the limit.

Next consider open string fields other than the tachyon.
The calculation for the tachyon
depends only on the two-point functions (\ref{disk-propagator})
and (\ref{cone-propagator}),
and two-point functions
of primary fields are uniquely determined
by their conformal dimensions.
Therefore, we conclude that
the kinetic terms
of the open string fields $\{ \varphi_i \}$
vanish when the corresponding vertex operator
${\cal O}_{\varphi_i (k)}$ is primary and has
conformal dimension one.
This condition is nothing but
the familiar physical state condition in string theory
so that the result in this subsection
provides strong evidence
that the correct mass spectrum of open string states
can be obtained in VSFT
based on our description
of the open string fields $\{ \varphi_i \}$.

Note that conditions
other than the on-shell condition
for a vertex operator to be physical,
such as the transversality condition
for the massless vector field, are also imposed
as we discussed at the end of Section 3.
The argument so far, however, does not guarantee
that the kinetic terms
of the open string fields $\{ \varphi_i \}$ are
correctly reproduced.
For example, it is not obvious that the kinetic term
for the massless gauge field takes a gauge-invariant
form. Any kinetic term of the form
\begin{equation}
  A_\mu (k) [ a(k^2) \eta^{\mu \nu} k^2
  + b(k^2) k^\mu k^\nu ] A_\nu (-k)
\end{equation}
vanishes when $k^2=0$ and $k \cdot A(k) =0$
for any pair of functions $a(k^2)$ and $b(k^2)$.
Gauge invariance requires that  $b(k^2)=-a(k^2)$.
It would be interesting to see if the gauge invariance
of the string field theory guarantees the gauge invariance
of the open string fields $\{ \varphi_i \}$.\footnote
{We can confirm, for example,
that $| \chi_{A_\mu (k)+k_\mu} \rangle$
and $| \chi_{A_\mu (k)} \rangle$ are gauge equivalent
when $k^2=0$ and $k \cdot A=0$
in the framework of Section 3,
namely, as far as we consider inner products
with wedge states which remain finite
in the large $n$ limit.
However, it is not clear if it holds in the framework of
the present section.}

There are other issues we have to address
regarding the argument in this subsection,
which we will discuss in the next section.

\subsection{Tachyon kinetic term}

In order to read off the on-shell three-tachyon coupling from
the cubic interaction, we need to normalize the tachyon field
canonically.
We have to calculate the coefficient in front of
$k^2-1$ in $K(k^2)$,
but the calculation is much more complicated
than that of $K(1)$
because of the conformal factor ${\cal F} (\theta)^{k^2-1}$
and the $k$-dependence in the propagator.
We present details of the calculation in Appendix B,
and explain an outline of the derivation in this subsection.

If we could set $\epsilon=0$,
the integrals of the vertex operators would be over
the whole boundary for each term
in $S[T(k)]$ (\ref{tachyon-action}).
In that case, $K(k^2)$ is given by
\begin{eqnarray}
  K(k^2) \bigg|_{\epsilon=0}
  &=& \frac{1}{2}
  \int_{0}^{2 \pi} d \theta_2 \int_{0}^{2 \pi} d \theta_1~
  {\cal F} (\theta_1)^{k^2-1}
  {\cal F} (\theta_2)^{k^2-1}
  \left| 2 \sin \frac{\theta_2-\theta_1}{2} \right|^{-2 k^2}
\nonumber \\
  && - \frac{1}{3}
  \int_{0}^{3 \pi} d \theta_2 \int_{0}^{3 \pi} d \theta_1~
  {\cal F} (\theta_1)^{k^2-1}
  {\cal F} (\theta_2)^{k^2-1}
  \left| 3 \sin \frac{\theta_2-\theta_1}{3} \right|^{-2 k^2},
\end{eqnarray}
where we have extended the definition of
${\cal F} (\theta)$ from
$-\pi/2 \le \theta \le \pi/2$ to all $\theta$ through
${\cal F} (\theta+\pi) = {\cal F} (\theta)$.
If we define $K(n, k^2)$ by
\begin{equation}
  K(n, k^2)
  \equiv \frac{1}{2 n}
  \int_{0}^{2 n \pi} d \theta_2
  \int_{0}^{2 n \pi} d \theta_1~
  {\cal F} (\theta_1)^{k^2-1}
  {\cal F} (\theta_2)^{k^2-1}
  \left| 2 n \sin \frac{\theta_2-\theta_1}{2 n}
  \right|^{-2 k^2},
\end{equation}
$K(k^2)$ is given by
\begin{equation}
  K(k^2) \bigg|_{\epsilon=0}
  = K(1, k^2) - K(3/2, k^2).
\label{K-from-K_n}
\end{equation}
Let us calculate $K(n, k^2)$.
Since we are interested in the region $k^2 \simeq 1$,
we expand the conformal factors around $k^2=1$ to find
\begin{eqnarray}
  && K(n, k^2)
  = \frac{1}{2 n}
  \int_{0}^{2 n \pi} d \theta_2
  \int_{0}^{2 n \pi} d \theta_1~
  \left| 2 n \sin \frac{\theta_2-\theta_1}{2 n}
  \right|^{-2 k^2}
\nonumber \\
  && \qquad \qquad \qquad \times
  \left\{ 1 + (k^2-1) \ln {\cal F}(\theta_1)
  + (k^2-1) \ln {\cal F} (\theta_2)
  + O( (k^2-1)^2 ) \right\}
\nonumber \\
  && = \frac{1}{2 n}
  \int_{0}^{2 n \pi} d \theta_2
  \int_{0}^{2 n \pi} d \theta_1~
  \left| 2 n \sin \frac{\theta_2-\theta_1}{2 n}
  \right|^{-2 k^2}
\nonumber \\
  && \qquad \qquad \times
  \left\{ 1 + 2 (k^2-1) \ln {\cal F}(\theta_2)
  + O( (k^2-1)^2 ) \right\}
\nonumber \\
  && = \frac{1}{2 n}
  \int_{0}^{2 n \pi} d \theta~
  \left| 2 n \sin \frac{\theta}{2 n}
  \right|^{-2 k^2}
  \int_{0}^{2 n \pi} d \theta'~
  \left\{ 1 + 2 (k^2-1) \ln {\cal F}(\theta') \right\}
  + O( (k^2-1)^2 )
\nonumber \\
  && =
  \int_{0}^{2 n \pi} d \theta~
  \left| 2 n \sin \frac{\theta}{2 n}
  \right|^{-2 k^2}
  \int_{-\pi/2}^{\pi/2} d \theta'~
  \left\{ 1 + 2 (k^2-1) \ln {\cal F}(\theta') \right\}
  + O( (k^2-1)^2 ),
\nonumber \\
\end{eqnarray}
where we used the periodicity of ${\cal F} (\theta)$
in the last step.
Note that the two integrals factorize in the last line
and the integral over $\theta'$ is independent of $n$.
The integral over $\theta$ does not converge near
the on-shell point $k^2 \simeq 1$.
That is why we needed
to introduce the regularization $\epsilon$.
In the momentum region $k^2 \simeq 1$,
the divergence is coming only from
the most singular part of the propagator
which is independent of details of the Riemann surface
and thus independent of $n$.
Therefore, the divergent part cancels when we compute
$K(k^2)$ through (\ref{K-from-K_n}).
Since the divergent part can be taken
to be analytic in $k^2$,
the finite part can be obtained by analytic continuation from
the region $\Re (k^2) < 1/2$
where the integral converges.\footnote{
I would like to thank Takuya Okuda
for the discussion on this point.
The explicit form of the singular part
when we use point-splitting regularization
is given in terms of the incomplete
beta function in (\ref{K-1}).}
It can be expressed
in terms of the beta function
as we see in (\ref{K-1})
and vanishes when $k^2=1$:
\begin{eqnarray}
  \int_{0}^{2 n \pi} d \theta~
  \left| 2 n \sin \frac{\theta}{2 n}
  \right|^{-2 k^2}
  &=& (2 n)^{-2 k^2+1}~
  B \left( \frac{1}{2}-k^2, \frac{1}{2} \right)
\nonumber \\
  &=& \frac{\pi}{n} (k^2-1) + O( (k^2-1)^2 ).
\end{eqnarray}
Therefore, $K(n, k^2)$ is given by
\begin{equation}
  K(n, k^2) = \frac{\pi^2}{n} (k^2-1) 
  + O( (k^2-1)^2 ),
\end{equation}
and $K(k^2)$ is
\begin{equation}
  K(k^2) = \frac{\pi^2}{3} (k^2-1) + O( (k^2-1)^2 ).
\label{K(k^2)-result}
\end{equation}
A more careful calculation
using point-splitting regularization
given in Appendix B
reproduces the same result (\ref{K(k^2)-result})
in the limit $\epsilon \to 0$
if ${\cal F} (\theta)$ is not too singular.
Note that the coefficient in front of $k^2-1$
is independent of the conformal factor ${\cal F} (\theta)$.
This is important for the consistency: since the on-shell
cubic interaction does not depend on ${\cal F} (\theta)$,
the on-shell three-tachyon coupling would depend on
${\cal F} (\theta)$
if this coefficient depended on ${\cal F} (\theta)$.
The tachyon field is therefore canonically normalized
as follows:
\begin{equation}
  \widehat{T} (k)
  = \left( \frac{{\cal K} \pi^2}{3} \right)^{1/2} T(k).
\label{normalized-tachyon}
\end{equation}

\subsection{Three-tachyon coupling and the D-brane tension}

The tachyon cubic term is given by
\begin{eqnarray}
  && S^{(3)} = - \vev{\Psi_g | {\cal Q} | \Psi_g}
  \bigg[ \vev{ {\it 3} | {\it 0} }
         + \vev{ {\it 2} | {\it 1} }
\nonumber \\
  && \qquad \qquad
         - \vev{ {\it 3} | {\it 0} \ast {\it 0} }
         - \vev{ {\it 2} | {\it 1} \ast {\it 0} }
         - \vev{ {\it 2} | {\it 0} \ast {\it 1} }
         - \frac{1}{3} \vev{ {\it 1} | {\it 1} \ast {\it 1} }
  \bigg].
\end{eqnarray}
This takes the following form:
\begin{eqnarray}
  S^{(3)} = -\frac{\cal K}{3} (2 \pi)^{26}
  \int d^{26} k_1~ d^{26} k_2~ d^{26} k_3~
  \delta (k_1+k_2+k_3)~ T(k_1) T(k_2) T(k_3)
\nonumber \\
  \times V(k_1, k_2, k_3).
\end{eqnarray}
We denote the on-shell value
of $V(k_1, k_2, k_3)$ by $V$,
which consists of the contributions from
$\vev{ {\it 3} | {\it 0} }$,
$\vev{ {\it 2} | {\it 1} }$,
$\vev{ {\it 3} | {\it 0} \ast {\it 0} }$,
$\vev{ {\it 2} | {\it 1} \ast {\it 0} }$,
$\vev{ {\it 2} | {\it 0} \ast {\it 1} }$,
and $\vev{ {\it 1} | {\it 1} \ast {\it 1} }$:
\begin{equation}
  - \frac{1}{3} V
  = - \left. \frac{1}{3} V(k_1, k_2, k_3)
    \right|_{k_1^2=k_2^2=k_3^2=1}
  = V_{30} + V_{21} - V_{300} - V_{210} - V_{201}
  - \frac{1}{3} V_{111}.
\label{V-definition}
\end{equation}
The contributions from
$\vev{ {\it 3} | {\it 0} }
- \vev{ {\it 3} | {\it 0} \ast {\it 0} }$
and
$\vev{ {\it 2} | {\it 1} }
- \vev{ {\it 2} | {\it 1} \ast {\it 0} }
- \vev{ {\it 2} | {\it 0} \ast {\it 1} }$
might be expected to vanish
if we naively use the equations of motion for
$\ket{\Xi_m}$ and $\ket{\chi_T (k)}$.
As can be anticipated from our experience
in Subsection 4.1, however, the calculations
using point-splitting regularization
presented in Appendix D show that they do not vanish
in the large $n$ limit:
\begin{equation}
  V_{30} - V_{300} \ne 0, \qquad
  V_{21} - V_{210} - V_{201} \ne 0.
\end{equation}
We find, however, a surprising cancellation between
the two expressions so that the sum turns out to vanish
in the large $n$ limit:
\begin{equation}
  V_{30} + V_{21} - V_{300} - V_{210} - V_{201} = o(\epsilon),
\label{V-cancellation}
\end{equation}
where we denote terms which vanish
in the limit $\epsilon \to 0$ by $o(\epsilon)$.
We will use this notation
throughout the rest of the paper.\footnote{
We distinguish $o(\epsilon)$ from $O(\epsilon)$.
The latter denotes terms of order $\epsilon$.
}
Therefore, only $V_{111}$ contributes to $V$:
\begin{equation}
    \frac{1}{3} V = \frac{1}{3} V_{111} + o(\epsilon).
\end{equation}
Let us calculate $V_{111}$. Since
\begin{equation}
  k_1 \cdot k_2
  = \frac{1}{2} (k_1 + k_2)^2
  - \frac{1}{2} k_1^2 - \frac{1}{2} k_2^2 = -\frac{1}{2},
\end{equation}
when $k_1^2=k_2^2=k_3^2=1$ and $k_1+k_2+k_3=0$,
and similarly $k_2 \cdot k_3 = k_3 \cdot k_1 = -1/2$,
$V_{111}$ is given by
\begin{eqnarray}
  && V_{111}
  = \int_{\pi/2+\epsilon/2}^{3\pi/2-\epsilon/2} d \theta_3
  \int_{-\pi/2+\epsilon/2}^{\pi/2-\epsilon/2} d \theta_2
  \int_{-3\pi/2+\epsilon/2}^{-\pi/2-\epsilon/2} d \theta_1
\nonumber \\
  && \qquad \qquad \times
  \left| 3 \sin \frac{\theta_2-\theta_1}{3} \right|^{-1}
  \left| 3 \sin \frac{\theta_3-\theta_1}{3} \right|^{-1}
  \left| 3 \sin \frac{\theta_3-\theta_2}{3} \right|^{-1}.
\label{V_111}
\end{eqnarray}
It turns out that $V_{111}$ is finite in the limit $\epsilon \to 0$
so that we can set $\epsilon$ to zero.
We present the calculation of the resulting integral in Appendix E
and the result is
\begin{equation}
  \int_{\pi/2}^{3\pi/2} d \theta_3
  \int_{-\pi/2}^{\pi/2} d \theta_2
  \int_{-3\pi/2}^{-\pi/2} d \theta_1
  \left| 3 \sin \frac{\theta_2-\theta_1}{3} \right|^{-1}
  \left| 3 \sin \frac{\theta_3-\theta_1}{3} \right|^{-1}
  \left| 3 \sin \frac{\theta_3-\theta_2}{3} \right|^{-1}
  = \frac{\pi^2}{3}.
\end{equation}
Therefore, we have
\begin{equation}
  V = \frac{\pi^2}{3} + o(\epsilon).
\label{V}
\end{equation}

The on-shell three-tachyon coupling $g_T$ is defined
by the on-shell value of the cubic interaction
when we express $S^{(3)}$ in terms of the canonically
normalized tachyon $\widehat{T}(k)$.
Namely, $g_T$ is given by
\begin{equation}
  g_T = \widehat{V} (k_1, k_2, k_3)
  \bigg|_{k_1^2=k_2^2=k_3^2=1},
\end{equation}
where $\widehat{V} (k_1, k_2, k_3)$ is defined by
\begin{eqnarray}
  S^{(3)} = -\frac{1}{3} (2 \pi)^{26}
  \int d^{26} k_1~ d^{26} k_2~ d^{26} k_3~
  \delta (k_1+k_2+k_3)~
  \widehat{T}(k_1) \widehat{T}(k_2) \widehat{T}(k_3)
\nonumber \\
  \times \widehat{V}(k_1, k_2, k_3).
\end{eqnarray}
  From the relation (\ref{normalized-tachyon})
between $T(k)$ and $\widehat{T}(k)$,
and the on-shell cubic interaction (\ref{V}),
$g_T$ is given by
\begin{equation}
  g_T = \left( \frac{{\cal K} \pi^2}{3} \right)^{-1/2}.
\end{equation}
  The tension of a single D25-brane $T_{25}$ predicted by
the three-tachyon coupling $g_T$ through (\ref{D25-brane-tension})
is given by
\begin{equation}
  T_{25} = \frac{1}{2 \pi^2 g_T^2}
  = \frac{\cal K}{6}.
\label{T=K/6}
\end{equation}
On the other hand, the energy density ${\cal E}_c$
of the classical solution $\ket{\Xi_m} \otimes \ket{\Psi_g}$
is given by
\begin{eqnarray}
  \int d^{26} x~ {\cal E}_c 
  &=& \frac{1}{2} \vev{\Xi_m | \Xi_m}
    \vev{\Psi_g | {\cal Q} | \Psi_g}
    + \frac{1}{3} \vev{\Xi_m | \Xi_m \ast \Xi_m}
    \vev{\Psi_g | \Psi_g \ast \Psi_g}
\nonumber \\
  &=& \frac{1}{6} \vev{\Xi_m | \Xi_m}
    \vev{\Psi_g | {\cal Q} | \Psi_g}
  = \int d^{26} x~ \frac{\cal K}{6}.
\end{eqnarray}
This is in perfect agreement with the interpretation
that the configuration $\ket{\Xi_m} \otimes \ket{\Psi_g}$
describes a single D25-brane:
\begin{equation}
  T_{25} = {\cal E}_c.
\end{equation}
This is our main result in this paper.

The calculations in Appendix D
are so complicated that it would be difficult
to confirm that the cancellation (\ref{V-cancellation})
does not depend on details of the regularization.
We will give a calculation in Subsection 5.4
which might be regarded as a piece of evidence
that the result (\ref{V}) is not
sensitive to details of the regularization.

\section{Discussion}
\setcounter{equation}{0}

In this section
we discuss some problematic issues
of our formulation.
We then discuss
a reformulation of our description
which is rather close to BSFT,
and end with future directions.

\subsection{Is the sliver state a classical solution?}

We found in Subsections 4.1 and 4.3
that the sliver state $\ket{\Xi_m}$
and the linearized on-shell open string states
$\ket{\chi_{\varphi_i} (k)}$
do not satisfy their equations of motion
when they are contracted with the class of states
$\ket{\{ \varphi_i \}}$ (\ref{general}),
\begin{eqnarray}
  \vev{ \{ \varphi_i \} | \Xi_m }
  &\ne& \vev{ \{ \varphi_i \} | \Xi_m \ast \Xi_m },
\nonumber \\
  \vev{ \{ \varphi_i \} | \chi_{\varphi_i} (k)}
  &\ne& \vev{ \{ \varphi_i \} | \chi_{\varphi_i} (k) \ast \Xi_m }
    + \vev{ \{ \varphi_i \} | \Xi_m \ast \chi_{\varphi_i} (k) },
\label{breakdown}
\end{eqnarray}
while they satisfy their equations of motion
when contracted
with an arbitrary state in the matter Fock space $\ket{\phi}$,
\begin{eqnarray}
  \vev{ \phi | \Xi_m }
  &=& \vev{ \phi | \Xi_m \ast \Xi_m },
\nonumber \\
  \vev{ \phi | \chi_{\varphi_i} (k)}
  &=& \vev{ \phi | \chi_{\varphi_i} (k) \ast \Xi_m }
    + \vev{ \phi | \Xi_m \ast \chi_{\varphi_i} (k) }.
\end{eqnarray}
Does this mean that the sliver state is not a classical
solution of VSFT
and that we are expanding the action
around an inappropriate configuration?

Let us see
if the linear terms
of the open string fields $\{ \varphi_i \}$
vanish when we express the VSFT action
in terms of them.
The part of the VSFT action
which is linear in $\{ \varphi_i \}$ is given by
\begin{equation}
  S^{(1)} = - \vev{\Psi_g | {\cal Q} | \Psi_g}
  \int d^{26} k~ \varphi_i (k)~
  \left[ 
  \vev{ \chi_{\varphi_i} (k) | \Xi_m }
  - \vev{ \chi_{\varphi_i} (k) | \Xi_m \ast \Xi_m }
  \right].
\end{equation}
The inner products 
$\vev{ \chi_{\varphi_i} (k) | \Xi_m }$
and $\vev{ \chi_{\varphi_i} (k) | \Xi_m \ast \Xi_m }$
are expressed in terms of one-point functions
of the vertex operator ${\cal O}_{\varphi_i (k)}$.
It is obvious from momentum conservation that
the one-point functions vanish for a nonzero momentum.
The vertex operators
for zero-momentum open string fields other than the tachyon
have a nonzero conformal dimension
so that their one-point functions vanish.
Therefore, the tachyon is
the only dangerous field which may have
a nonvanishing linear term.

Let us therefore calculate the tachyon potential $V(T)$
in our description of the tachyon field
$| e^{-T} \rangle$.
For a constant tachyon field $T(x)=T$,
the factor inserted in $| e^{-T} \rangle$
in the cone representation is given by
\begin{equation}
  \exp \left[ -T
  \int_{-(n-1) \pi/2}^{(n-1) \pi/2} d \theta~
  {\cal F}_0 (\theta)^{-1} \right]
  = e^{ -a T},
\end{equation}
where the value of a constant $a$,
\begin{equation}
  a = \int_{-(n-1) \pi/2}^{(n-1) \pi/2} d \theta~
  {\cal F}_0 (\theta)^{-1},
\end{equation}
depends on the choice of the Riemann surface $\Sigma_n$
where the off-shell tachyon is defined.
The tachyon potential $V(T)$ is easily calculated from
(\ref{tachyon-action}) and the result is
\begin{equation}
  V(T) = {\cal K} \left( \frac{1}{2} e^{-2 a T}
         - \frac{1}{3} e^{-3 a T} \right)
       = 3~ T_{25}~ e^{-2 a T}
         - 2~ T_{25}~ e^{-3 a T},
\label{tachyon-potential}
\end{equation}
where we used (\ref{T=K/6}).
The shape of $V(T)$ is given in Figure 1.
\begin{figure}
\centerline{\epsfxsize=4in\epsfbox{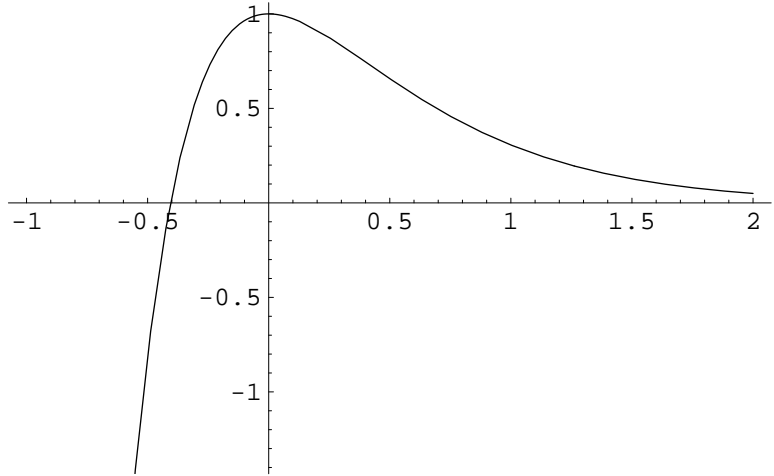}}
\hspace{1cm}
\medskip
\caption{The tachyon potential $V(T)$.
The horizontal axis is $aT$
and the vertical axis is $V(T)/T_{25}$.
The sliver state corresponds to $T=0$
and the tachyon vacuum corresponds to
$T=\infty$.}
\end{figure}
The linear term vanishes at $T=0$,
\begin{equation}
  \left. \frac{d V(T)}{d T} \right|_{T=0} = 0.
\end{equation}
The tachyon vacuum corresponds to $T=\infty$
where $V(T)=0$,
and the potential height at $T=0$ is exactly
the same as the D25-brane tension as we calculated
in Subsection 4.3:
\begin{equation}
  V(0) - V(\infty) = T_{25}.
\end{equation}

To summarize, the linear terms of the open string fields,
including the tachyon, vanish at the configuration
where $\varphi_i = 0$,
which corresponds to the sliver state:
\begin{equation}
  \left. \frac{\delta S[\{ \varphi_i \}]}
              {\delta \varphi_i} \right|_{\varphi_i=0}
  = 0.
\end{equation}
In this sense, we can regard the sliver state
as a classical solution.
However, we do not completely understand
whether the breakdown of the equations of motion
in the form of (\ref{breakdown}) is problematic or not.

\subsection{The large $n$ limit and renormalization}

The primary goal of the present paper was
to study if VSFT describes
the ordinary perturbative dynamics
of open strings based on our proposal.
For this purpose, we calculated the kinetic term
of the tachyon near its mass shell and the on-shell
three-tachyon coupling in Section 4.
We found that all the divergences we encountered
in these calculations canceled
and these quantities stay finite
in the large $n$ limit.
However, the tachyon state $| e^{-T} \rangle$ itself,
or more general states $\ket{\{ \varphi_i \}}$, do not
seem to have the large $n$ limit.

As we have seen in Section 4,
singularities in the large $n$ limit
become short-distance singularities
such as $1/\epsilon$ or $\ln \epsilon$
when we calculate correlation functions
on the unit disk or the cone with an angle $3 \pi$.
The open string fields $\{ \varphi_i \}$
correspond to bare coupling constants
in the open string sigma model and
we may need to renormalize them
appropriately in the large $n$
limit to make our description more well-defined.

It might be useful to notice that there were
two different types of divergences in the calculations
in Section 4. Take $\vev{ {\it 2} | {\it 0} }$
as an example. The $1/\epsilon$ divergence of $K_{20}(1)$
in (\ref{on-shell-K_20}) comes from the contribution
where the two vertex operators become close in the bulk
of the integration region.
On the other hand,
the $\ln \epsilon$ divergence of $K_{20}(1)$
in (\ref{on-shell-K_20}) can be regarded as
a boundary effect of the integration region.
To understand this, the following analogy might be
helpful.
The inner product $\vev{ {\it 2} | {\it 0} }$
is the quadratic part of $\vev{ \{ \varphi_i \} | \Xi_m }$
in the expansion of $T(k)$
when we turn on only the tachyon field.
The boundary interaction
\begin{equation}
  \exp \left[ - \int dt \int d^{26} k
  \sum_i \varphi_i (k)
  {\cal O}_{\varphi_i (k)} (t) \right]
\end{equation}
in (\ref{general}) is introduced only on a part of the boundary
in  $\vev{ \{ \varphi_i \} | \Xi_m }$.
The inner product $\vev{ \{ \varphi_i \} | \Xi_m }$
is therefore analogous to an open Wilson line
in ordinary gauge theory in this respect.
On the other hand, the boundary interaction
is introduced on the whole boundary
in the inner products
$\vev{ \{ \varphi_i \} | \{ \varphi_i \} }$ and
$\vev{ \{ \varphi_i \} | \{ \varphi_i \} \ast \{ \varphi_i \} }$
appearing in the VSFT action (\ref{varphi-action}).
These inner products are analogous
to closed Wilson loops.
When we expand the Wilson-loop-like inner products
$\vev{ \{ \varphi_i \} | \{ \varphi_i \} }$ and
$\vev{ \{ \varphi_i \} | \{ \varphi_i \} \ast \{ \varphi_i \} }$
as we did in (\ref{S^2}),
the combinations
$\vev{ {\it 1} | {\it 1} }/2 + \vev{ {\it 2} | {\it 0} }$
and $\vev{ {\it 1} | {\it 1} \ast {\it 0} }
+ \vev{ {\it 2} | {\it 0} \ast {\it 0} }$ 
appear, and the $\ln \epsilon$ divergence is absent from
$K_{11}(1)/2 + K_{20}(1)$
and $K_{110}(1) + K_{200}(1)$
corresponding to these combinations of inner products.\footnote
{The absence of the $\ln \epsilon$ divergence
is more transparent in the calculations of
$\widetilde{K}_{11}/2 + \widetilde{K}_{20}$
and $\widetilde{K}_{110} + \widetilde{K}_{200}$
in Appendix C.}
We can generally expect that the divergence
coming from the boundary of the boundary interaction
cancels in calculations of Wilson-loop-like
quantities such as the VSFT action.

The other type of divergence such as $1/\epsilon$
in $K_{20}(1)$ is familiar in the open string sigma model.
We will be able to handle such divergences
by a conventional renormalization procedure,
at least for renormalizable boundary interactions,
and the situation is similar to that of BSFT.
Note also that a multiplicative renormalization
of the tachyon field will not change the on-shell
three-tachyon coupling constant $g_T$ we calculated
in Section 4.

On the other hand,
we do not know how to handle
the divergence coming from the boundary
of the boundary interaction.
If this class of divergence remains
in a physically relevant calculation,
it can be a problem of our formulation,
although we have not encountered such situations so far
and we do not expect such problems
in the calculation of the VSFT action
as we mentioned before.
Incidentally, inner products of general states
with a state in the matter Fock space,
$\vev{ \{ \varphi_i \} | \phi}$,
become Wilson-loop-like in the large $n$ limit.
It seems therefore possible to make them well-defined
by a conventional renormalization procedure
for renormalizable interactions,
although it is not clear
if the inner products $\vev{ \{ \varphi_i \} | \phi}$
are really physically relevant in VSFT.

\subsection{Off-shell definitions}

In Section 2 we mentioned an ambiguity
in our off-shell definition of open string fields.
If we need to renormalize the open string fields
as we discussed in the previous subsection,
the choice of the renormalization scheme
will be another source of off-shell ambiguity.\footnote
{These two ambiguities might not be independent.
The conformal factor ${\cal F}_0 (\theta)^{k^2-1}$
in (\ref{cone-integral}) has an implicit dependence
on $n$ through the choice of the Riemann surface $\Sigma_n$,
and the $\theta$-independent part of this factor
looks similar to a multiplicative renormalization of $T(k)$.}
It is important for the consistency of our formulation
that such off-shell ambiguity does not
affect physically relevant quantities.
We found, for example, that
the relation between the D25-brane tension $T_{25}$
and the on-shell three-tachyon coupling constant $g_T$
derived in Section 4 was independent of
the ambiguity coming from ${\cal F}_0 (\theta)$
in (\ref{cone-integral}).
If the physics is really independent
of the off-shell ambiguity,
we can in principle choose any off-shell definition.
However, an inappropriate choice might cause
a singular behavior of the off-shell fields.

If we take the Riemann surface
$\Sigma_n$ to be a cone with an angle $n \pi$,
which corresponds to ${\cal F}_0 (\theta) = 1$
in (\ref{cone-integral}) for the tachyon,
the tachyon kinetic term can be calculated exactly
for $\Re (k^2) < 1/2$.
  From the calculation in Subsection 4.2, it is given by
\begin{eqnarray}
  \frac{1}{2} K(k^2)
  &=& \frac{\pi}{2}
  \left( \frac{1}{n-1} \right)^{2 k^2 -2}
  ( 2^{-2 k^2+1} - 3^{-2 k^2+1} )~
  B \left( \frac{1}{2}-k^2, \frac{1}{2} \right)
\nonumber \\
  &=& \frac{\pi}{2}
  \left( \frac{1}{n-1} \right)^{2 k^2 -2}
  ( 2^{-2 k^2+1} - 3^{-2 k^2+1} )~
  \frac{\sqrt{\pi}~
  \Gamma \left( \frac{1}{2}-k^2 \right)}
  {\Gamma (1-k^2)}.
\label{all-k}
\end{eqnarray}
We may renormalize the tachyon field
to absorb the $n$-dependent factor in (\ref{all-k})
coming from (\ref{F_0-to-F}), which is not relevant
to the present discussion.
If we assume the analyticity in $k^2$,
we can define $K(k^2)$ for all $k^2$ by analytic continuation.
We then note that
the kinetic term (\ref{all-k}) vanishes not only at $k^2=1$
but also at any positive integer $k^2= 1,2,3, \ldots$.
The kinetic term is also singular
at $k^2 = 3/2, 5/2, \ldots,$ where it diverges.
If this implies the existence of an infinite number of
tachyons, the theory will definitely be pathological,
and it can be a problem of our formulation.

We have argued that it is universal that
the tachyon kinetic term vanishes when $k^2=1$,
but it is not clear if other zeros of (\ref{all-k})
at $k^2=2,3, \ldots$ are also universal.
The singular behavior of (\ref{all-k})
for higher $k^2$ might be an artifact
of an inappropriate off-shell definition.
In fact, when we increase the momentum $k$ in the calculation
of $K(k^2)$ using point-splitting regularization,
the next-to-leading singularity in (\ref{n-independence})
becomes divergent at $k^2=3/2$,
and it is $n$-dependent.
The divergence in $K(k^2)$ will no longer cancel
and the tachyon field beyond $k^2=3/2$
seems to depend strongly on the off-shell ambiguity.
It is not clear if the tachyon kinetic term
universally vanishes at $k^2=2,3, \ldots$.
We cannot claim anything definite about this issue
for now,
but we hope that the expression (\ref{all-k})
corresponds to a singular definition of the tachyon and
there is a better class of
off-shell definitions.

\subsection{BSFT-like reformulation}

So far we have found that
when we express the VSFT action
in terms of the open string fields $\{ \varphi_i \}$
based on our proposal (\ref{general}),
\begin{enumerate}
\item
the linear terms vanish,
\item
the kinetic terms vanish
when the fields $\{ \varphi_i \}$ satisfy
the physical state conditions,
\item
and the relation between the D25-brane tension $T_{25}$
and the on-shell three-tachyon coupling $g_T$
given by (\ref{D25-brane-tension})
is correctly reproduced.
\end{enumerate}
Although we started from VSFT,
most of the calculations in the present paper
are reminiscent of
those of boundary string field theory (BSFT)
\cite{Witten:1992qy, Witten:1992cr, Li:1993za,
Shatashvili:1993kk, Shatashvili:1993ps,
Gerasimov:2000zp, Kutasov:2000qp}\footnote{
In particular, similar calculations can be found
in Appendix A of \cite{Kutasov:2000qp}.
Related calculations can also be found
in a different context in \cite{Recknagel:1998ih}.
}.
Furthermore,
the problems we discussed in Subsections
5.1 and 5.2 seem to suggest a BSFT-like
reformulation of our description.
In fact, all the results we just mentioned
are effectively reproduced
by the following BSFT-like action:
\begin{eqnarray}
  S[ \{ \varphi_i \} ]
  &=& - 3~ T_{25} \vev{
  \exp \left[
  - \int d^{26} k \sum_i \varphi_i (k)
  \int_{0}^{2 \pi} d \theta~
  {\cal O}_{\varphi_i (k)} (e^{i \theta})
  \right] }_{\rm disk}
\nonumber \\
  && + 2~ T_{25} \vev{
  \exp \left[
  - \int d^{26} k \sum_i \varphi_i (k)
  \int_{0}^{3 \pi} d \theta~
  {\cal O}_{\varphi_i (k)} (e^{i \theta})
  \right] }_{3 \pi},
\label{BSFT}
\end{eqnarray}
where the correlation functions are normalized as
\begin{equation}
  \vev{1}_{\rm disk} = \int d^{26} x, \quad
  \vev{1}_{3 \pi} = \int d^{26} x,
\end{equation}
and we assume an appropriate regularization and renormalization
scheme.
As we have discussed,
the off-shell definition for $\{ \varphi_i \}$ is not unique.
For example, we can define off-shell tachyon field
taking the ambiguity into account as follows:
\begin{eqnarray}
  S[ T(k) ]
  &=& - 3~ T_{25} \vev{
  \exp \left[
  - \int d^{26} k~ T(k)
  \int_{0}^{2 \pi} d \theta~ {\cal F} (\theta)^{k^2-1}
  e^{ikX} (e^{i \theta})
  \right] }_{\rm disk}
\nonumber \\
  && + 2~ T_{25} \vev{
  \exp \left[
  - \int d^{26} k~ T(k)
  \int_{0}^{3 \pi} d \theta~ {\cal F} (\theta)^{k^2-1}
  e^{ikX} (e^{i \theta})
  \right] }_{3 \pi},
\end{eqnarray}
where we extended the definition of
${\cal F} (\theta)$ from
$-\pi/2 \le \theta \le \pi/2$ to all $\theta$ through
${\cal F} (\theta+\pi) = {\cal F} (\theta)$
as we did in Subsection 4.2.
We can easily see that
the linear terms of $\{ \varphi_i \}$ vanish in the action
(\ref{BSFT}) and the tachyon potential is given by
(\ref{tachyon-potential}).
If we use point-splitting regularization,
we can also show that
the kinetic terms for $\{ \varphi_i \}$ vanish when
${\cal O}_{\varphi_i (k)}$ is primary
with conformal dimension one,
and the calculations for the tachyon kinetic term
near its mass shell are simpler than
those in Subsection 4.2 and Appendix B,
and give the same result.

The calculation for
the on-shell three-tachyon coupling
is remarkably simpler than the VSFT calculation
of Appendices D and E.
Before regularizing it,
the on-shell cubic interaction $V$
defined by (\ref{V-definition}) can be written as follows:
\begin{equation}
  - \frac{1}{3} V = V(1) - V(3/2),
\end{equation}
where $V(n)$ is defined by
\begin{eqnarray}
  && V(n)
  \equiv \frac{1}{2 n} \frac{1}{3!}
  \int_{0}^{2 n \pi} d \theta_3
  \int_{0}^{2 n \pi} d \theta_2
  \int_{0}^{2 n \pi} d \theta_1
\nonumber \\
  && \qquad \qquad \times
  \left| 2 n \sin \frac{\theta_1 - \theta_2}{2 n} \right|^{-1}
  \left| 2 n \sin \frac{\theta_2 - \theta_3}{2 n} \right|^{-1}
  \left| 2 n \sin \frac{\theta_3 - \theta_1}{2 n} \right|^{-1}
\nonumber \\
  && = \frac{\pi}{6}
  \int_{0}^{2 n \pi} d \theta_2
  \int_{0}^{2 n \pi} d \theta_1
  \left| 2 n \sin \frac{\theta_1 - \theta_2}{2 n} \right|^{-1}
  \left| 2 n \sin \frac{\theta_1}{2 n} \right|^{-1}
  \left| 2 n \sin \frac{\theta_2}{2 n} \right|^{-1}.
\end{eqnarray}
If we use point-splitting regularization,
$V(n)$ is regularized and calculated in the following way:
\begin{eqnarray}
  && V(n)
  = \frac{\pi}{3}
  \int_{2 \epsilon}^{2 n \pi -\epsilon} d \theta_2
  \int_{\epsilon}^{\theta_2 -\epsilon} d \theta_1
  \left( 2 n \sin \frac{\theta_2 - \theta_1}{2 n} \right)^{-1}
  \left( 2 n \sin \frac{\theta_1}{2 n} \right)^{-1}
  \left( 2 n \sin \frac{\theta_2}{2 n} \right)^{-1}
\nonumber \\
  && = \frac{2 \pi}{3 (2 n)^2}
  \int_{2 \epsilon}^{2 n \pi -\epsilon} d \theta_2
  \left( \sin \frac{\theta_2}{2 n} \right)^{-2}
  \ln \frac{\sin \frac{\theta_2 - \epsilon}{2 n}}
           {\sin \frac{\epsilon}{2 n}}
  = -\frac{\pi^2}{3 n}
    \left( 1 - \frac{3 \epsilon}{2 n \pi} \right)
  + \frac{\pi}{n} \cot \frac{\epsilon}{2 n}
    \ln \frac{\sin \frac{\epsilon}{n}}
             {\sin \frac{\epsilon}{2 n}}
\nonumber \\
  && = \frac{2 \pi \ln 2}{\epsilon}
    -\frac{\pi^2}{3 n} + O(\epsilon).
\end{eqnarray}
Therefore, $V$ is given by
\begin{equation}
  \frac{1}{3} V = \frac{\pi^2}{9} + O(\epsilon).
\end{equation}
This coincides with the result (\ref{V}),
and therefore gives
the relation between $T_{25}$
and $g_T$ (\ref{D25-brane-tension}) correctly.
This calculation seems to indicate that
the complication in the calculations of Appendix D
is due to the existence of the boundary of the boundary
interaction and the final result for $V$, (\ref{V}),
is not sensitive to details of the regularization.

Since this BSFT-like formulation is simpler
than VSFT for this kind of calculation,
it might be useful to test
if other aspects of string perturbation theory
are correctly reproduced.
It would also be interesting to explore
if the action (\ref{BSFT}) itself defines
a new string field theory.
The problem
regarding nonrenormalizable
boundary interactions
would be taken over from that of ordinary BSFT.
However, we expect some cancelation of the divergences
between the two terms in (\ref{BSFT})
as we have seen in Section 4 and Appendices B and D.
It would be important to understand
the structure of the divergences in (\ref{BSFT}),
or in (\ref{varphi-action}).

Interestingly, the results we listed at the beginning
of this subsection are also
reproduced by a class of BSFT-like actions\footnote{
I would like to thank Takuya Okuda for suggesting
this possibility.
} such as
\begin{eqnarray}
  S_{n,m} [ \{ \varphi_i \} ]
  &=& - \frac{\cal K}{2 n} \vev{
  \exp \left[
  - \int d^{26} k \sum_i \varphi_i (k)
  \int_{0}^{2 n \pi} d \theta~
  {\cal O}_{\varphi_i (k)} (e^{i \theta})
  \right] }_{2 n \pi}
\nonumber \\
  && + \frac{\cal K}{2 m} \vev{
  \exp \left[
  - \int d^{26} k \sum_i \varphi_i (k)
  \int_{0}^{2 m \pi} d \theta~
  {\cal O}_{\varphi_i (k)} (e^{i \theta})
  \right] }_{2 m \pi},
\label{generalized-BSFT}
\end{eqnarray}
with $n < m$,\footnote{
The ratio $n/m$ must be rational
if we want to incorporate
the off-shell ambiguity we mentioned in Section 2.
}
or a linear combination of $S_{n,m}$'s.
The action (\ref{BSFT}) corresponds to
the case where $n/m=2/3$.
It would be important to understand
whether this value, which is inherited from
Witten's open string field theory \cite{Witten:1985cc},
has any special meaning or not.

If we take the large $m$ limit
while $n$ is kept finite ($n/m \to 0$)
after point-splitting regularization,
the second term in (\ref{generalized-BSFT})
just subtracts the divergences
and does not contribute to the finite terms
as far as the calculations we have done so far
are concerned.
In this limit, the action may be related
to the (renormalized)
partition function with the boundary interaction
\cite{Fradkin:ys, Fradkin:1984pq, Fradkin:1985qd,
Abouelsaood:gd, Callan:1986bc}.\footnote{
See also a recent work \cite{Tseytlin:2000mt}.
}
The tachyon potential, however, becomes singular
in this limit.

If we choose $m=n+a$ and take $n$ to be large
while $a$ is kept finite ($n/m \to 1$),
the two terms in (\ref{generalized-BSFT}) become
almost the same. A conformal transformation which maps
one Riemann surface to the other becomes infinitesimal
so that the resulting action may be related
to the BSFT \cite{Witten:1992qy}
where the anticommutator of the BRST charge
with the vertex operator is inserted.
Let us calculate the tachyon potential in this limit.
Under an appropriate normalization
for the constant tachyon field $T$, it is given by
\begin{equation}
  V(T) = \frac{\cal K}{2 n} e^{-n T}
         - \frac{\cal K}{2 (n+a)} e^{-(n+a) T}.
\end{equation}
Since the D25-brane tension $T_{25}$ is given by
\begin{equation}
  T_{25} = \frac{\cal K}{2 n} - \frac{\cal K}{2 (n+a)}
         = \frac{{\cal K} a}{2 n (n+a)},
\end{equation}
the tachyon potential is normalized as follows:
\begin{equation}
  \frac{V(T)}{T_{25}} = \frac{n+a}{a} e^{-n T}
                      - \frac{n}{a} e^{-(n+a) T}.
\end{equation}
We can take the large $n$ limit if we redefine
the tachyon as $T \to T/n$ and the resulting potential
is given by
\begin{equation}
  \frac{V(T)}{T_{25}} = (T+1) e^{-T}.
\end{equation}
This coincides with the tachyon potential in BSFT
\cite{Witten:1992cr, Gerasimov:2000zp, Kutasov:2000qp}.
It would be interesting to learn more about
the relation between
the action (\ref{generalized-BSFT}) and BSFT.

\subsection{Ghost solution}

We have concentrated on the matter part of VSFT
assuming the existence of the universal
solution in the ghost part.
Actually, we implicitly assume more about
the ghost solution.

In the CFT formulation of string field theory
\cite{LeClair:1988sp},
we implicitly use the generalized gluing
and resmoothing theorem \cite{LeClair:1988sj}.
As is emphasized in \cite{Asakawa:1998dv},
the theorem holds only when the total central
charge vanishes.
Furthermore, we have to make the same conformal
transformation for both matter and ghost sectors.
Otherwise a conformal anomaly effectively occurs
even when the total central charge vanishes.

Since all the states we considered in this paper
were defined as the large $n$ limit of wedge states
with some operator insertions, the universal ghost
solution also has to share this property.
Namely, we assume the existence of a series of
purely ghost operators ${\cal Q} (n)$
and ghost states $\ket{\Psi_g (n)}$ labeled by $n$
satisfying the following conditions:
\begin{enumerate}
\item
the ghost operators ${\cal Q} (n)$
have vanishing cohomology,
at least in the large $n$ limit,
\item
the ghost states $\ket{\Psi_g (n)}$
take the form of wedge states labeled by $n$
with some ghost insertions,
\item
they solve the ghost equation of motion
in the large $n$ limit:
\begin{equation}
  \lim_{n \to \infty}
  \vev{ \Psi_g (n) | {\cal Q} (n) | \Psi_g (n) }
  + \lim_{n \to \infty}
  \vev{ \Psi_g (n) | \Psi_g (n) \ast \Psi_g (n) } = 0,
\label{ghost-condition-1}
\end{equation}
\item
and they give a finite D-brane tension:
\begin{equation}
  \lim_{n \to \infty} \vev{ n | n }
  \vev{ \Psi_g (n) | {\cal Q} (n) | \Psi_g (n) }
  = \int d^{26} x~ {\cal K},
\label{ghost-condition-2}
\end{equation}
where ${\cal K}$ is finite.
\end{enumerate}

Can we find such operators ${\cal Q} (n)$
and states $\ket{\Psi_g (n)}$?
The ghost solution found
by Hata and Kawano \cite{Hata:2001sq}
turned out to be described as the sliver state
of the twisted ghost CFT \cite{Gaiotto:2001ji}.
Unfortunately, the total central charge does not vanish
when we twist the ghost CFT.
However, we can show that a class of wedge states
in the twisted ghost CFT are described
by wedge states in the untwisted ghost CFT
with some operator insertions \cite{Okawa}.
We assume that such ghost states
satisfy the conditions (\ref{ghost-condition-1})
and (\ref{ghost-condition-2})
under an appropriate regularization.
We hope to report on this issue
in a future work \cite{Okawa}.

\subsection{Future directions}

In this paper we used the CFT formulation
\cite{LeClair:1988sp, LeClair:1988sj}
of string field theory.
In the development of VSFT
the interplay among various formulations
has played an important role.
The D25-brane solution in VSFT
was first studied in the operator formalism
\cite{Kostelecky:2000hz, Hata:2001sq}.
The technology in the operator formalism
of string field theory
\cite{Gross:1986ia, Gross:1986fk,
Cremmer:1986if, Samuel:1986qk, Ohta:wn}
is developing rapidly
based on the spectroscopy
of the Neumann coefficient matrices
\cite{Moore:2002fg, Rastelli:2001hh, Okuyama:2002yr,
Feng:2002rm, Feng:2002ib, Chen:2002md}.
It is used for analytical proofs
of conjectured equivalence
between results from the CFT descriptions
and corresponding ones in the operator formalism
\cite{Okuyama:2002yr, Okuyama:2002tw,
Okuda:2002fj, Bonora:2002iq}.
The half-string or split-string picture
\cite{Rastelli:2001rj, Gross:2001rk,
Kawano:2001fn, Gross:2001yk}
is particularly useful
when we consider systems of multiple D-branes.
It is further related to a recent reformulation
in terms of noncommutative field theory
\cite{Bars:2001ag, Douglas:2002jm}.
It would be important to study
relations of our description
to these other formulations
for future investigations on
generalizations to non-Abelian
cases, the role of the gauge invariance, and so on.
An approach to non-Abelian structures from
the viewpoint of BSFT such as \cite{Gerasimov:2001pg}
may also be useful.

We used wedge states to formulate our description
of the open string fields on a D25-brane
by making use of
their simple star algebra \cite{Rastelli:2000iu}.
On the other hand, other star algebra projectors
such as the butterfly state were also
found and studied \cite{Schnabl:2002ff, Gaiotto:2002kf}.
It would be interesting to consider
the description of open string fields
for other surface states
and see if the results such as the ones
we listed at the beginning of Subsection 5.4
are independent of the choice of the surfaces.

Finally, it would be
very important to understand
relations of our formulation
to Witten's cubic string field theory \cite{Witten:1985cc}
as well as further relations to BSFT \cite{Witten:1992qy}.
Witten's cubic string field theory
is based on the BRST quantization
while our formulation is reminiscent of
the old covariant quantization.
We hope that our formulation will give
some insights into dictionaries
between Witten's cubic string field theory,
BSFT, and VSFT.

\section*{Acknowledgements}
I would like to thank Teruhiko Kawano and Takuya Okuda
for collaborative discussions throughout the work
and Hirosi Ooguri for useful discussions.
I also thank Martin Schnabl for his help
in the analytic evaluation of the integral $I_{3/2}$
in Appendix D.
This work was supported in part by
the DOE grant DE-FG03-92ER40701
and by a McCone Fellowship in Theoretical Physics from
California Institute of Technology.


\newpage
\appendix
\renewcommand{\thesection}{Appendix \Alph{section}.}
\renewcommand{\theequation}{\Alph{section}.\arabic{equation}}

\section{The relation between $T_{25}$ and $g_T$}
\setcounter{equation}{0}

The inverse of the D25-brane tension $T_{25}$ and
the square of the on-shell three-tachyon coupling constant
$g_T$ are both proportional to the string coupling constant.
The dimensionless quantity $\alpha'^3 \, T_{25} \, g_T^2$
is therefore independent of the string coupling constant.
Let us calculate it
following the convention of \cite{Polchinski:rq}.

The effective action for the tachyon is given
in (6.5.16) of \cite{Polchinski:rq} by
\begin{equation}
  S = \frac{1}{{g'_o}^2} \int d^{26} x
  \left[ -\frac{1}{2} \partial_\mu T(x) \partial^\mu T(x)
         +\frac{1}{2 \alpha'} T(x)^2
         +\frac{1}{3} \sqrt{\frac{2}{\alpha'}} T(x)^3 \right].
\label{Polchinski-action}
\end{equation}
The tension of a D$p$-brane $T_p$ is given
in (8.7.26) of \cite{Polchinski:rq} by
\begin{equation}
  T_p^2 = \frac{\pi}{256 \kappa^2} ( 4 \pi^2 \alpha' )^{11-p}.
\end{equation}
The relation between $g'_o$ and $\kappa$ is given
in (8.7.28) of \cite{Polchinski:rq} by
\begin{equation}
  \frac{4 \pi \alpha' {g'_o}^2 }{\kappa}
  = 2^{18} \pi^{25/2} \alpha'^6.
\end{equation}
Therefore, the D25-brane tension $T_{25}$ is expressed
in terms of $g'_o$ by
\begin{equation}
  T_{25} = \frac{1}{4 \pi^2 \alpha'^2 {g'_o}^2 }.
\end{equation}
The on-shell three-tachyon coupling $g_T$ is defined
by\footnote{
The normalized tachyon $\widehat{T} (x)$ here
is related to $\widehat{T} (k)$ in Subsection 4.2 as
$$
  \widehat{T}_{\rm Polchinski} (x)
  = - \widehat{T}_{\rm ours} (x)
$$
with
$$
  \widehat{T}_{\rm ours} (x)
  = \int d^{26} k~ \widehat{T} (k) e^{ikX}.
$$
}
\begin{equation}
  S = \int d^{26} x
  \left[ -\frac{1}{2} \partial_\mu \widehat{T} (x)
                      \partial^\mu \widehat{T} (x)
         +\frac{1}{2 \alpha'} \widehat{T} (x)^2
         +\frac{1}{3} g_T \widehat{T} (x)^3 \right].
\end{equation}
Since the normalized tachyon field $\widehat{T}$ is related
to $T$ in (\ref{Polchinski-action}) as
\begin{equation}
  \widehat{T} (x) = \frac{T(x)}{g'_o},
\end{equation}
the relation between $g_T$ and $g'_o$ is given by
\begin{equation}
  g_T = g'_o \sqrt{\frac{2}{\alpha'}}.
\end{equation}
The D25-brane tension $T_{25}$ is therefore expressed
in terms of $g_T$ as
\begin{equation}
    T_{25} = \frac{1}{2 \pi^2 \alpha'^3 g_T^2 }.
\end{equation}

\section{$K(k^2)$}
\setcounter{equation}{0}

We calculate $K(k^2)$ in (\ref{K-from-4-terms})
when $k^2$ is nearly on shell $k^2 \simeq 1$
to show (\ref{K(k^2)-result})
in this appendix.
Let us begin with $K_{20}(k^2)$ (\ref{K_20})
and $K_{200}(k^2)$ (\ref{K_200}).
Both take the following form:
\begin{equation}
  K_{20/200}(n, k^2) =
  \int_{-\pi/2+3\epsilon/2}^{\pi/2-\epsilon/2} d \theta_2
  \int_{-\pi/2+\epsilon/2}^{\theta_2-\epsilon} d \theta_1~
  {\cal F}(\theta_1)^{k^2-1} {\cal F} (\theta_2)^{k^2-1}
  \left| 2 n \sin \frac{\theta_2 - \theta_1}{2 n}
  \right|^{-2 k^2},
\label{K_20/200}
\end{equation}
with $n=1$ for $K_{20}$ and $n=3/2$ for $K_{200}$.
Using the formula
\begin{equation}
  \int_{-\pi/2+3\epsilon/2}^{\pi/2-\epsilon/2} d \theta_2
  \int_{-\pi/2+\epsilon/2}^{\theta_2-\epsilon} d \theta_1~
  f(\theta_1,\theta_2)
  = \int_{-\pi/2+\epsilon/2}^{\pi/2-3\epsilon/2} d \theta_2
  \int_{\theta_2+\epsilon}^{\pi/2-\epsilon/2} d \theta_1~
  f(\theta_2,\theta_1)
\end{equation}
for any function $f$ with two variables,
$K_{20/200}(n, k^2)$ is rewritten in the following way:
\begin{eqnarray}
  && K_{20/200}(n, k^2)
\nonumber \\ 
 &=& \frac{1}{2}
  \int_{-\pi/2+3\epsilon/2}^{\pi/2-\epsilon/2} d \theta_2
  \int_{-\pi/2+\epsilon/2}^{\theta_2-\epsilon} d \theta_1~
  {\cal F}(\theta_1)^{k^2-1} {\cal F} (\theta_2)^{k^2-1}
  \left| 2 n \sin \frac{\theta_2 - \theta_1}{2 n}
  \right|^{-2 k^2}
\nonumber \\
  && + \frac{1}{2}
  \int_{-\pi/2+\epsilon/2}^{\pi/2-3\epsilon/2} d \theta_2
  \int_{\theta_2+\epsilon}^{\pi/2-\epsilon/2} d \theta_1~
  {\cal F}(\theta_1)^{k^2-1} {\cal F} (\theta_2)^{k^2-1}
  \left| 2 n \sin \frac{\theta_2 - \theta_1}{2 n}
  \right|^{-2 k^2}
\nonumber \\
  &=& \frac{1}{2}
  \int_{-\pi/2+3\epsilon/2}^{\pi/2-\epsilon/2} d \theta_2
  \int_{-\pi/2+\epsilon/2}^{\theta_2-\epsilon} d \theta_1~
  \left| 2 n \sin \frac{\theta_2 - \theta_1}{2 n}
  \right|^{-2 k^2}
\nonumber \\
  && \qquad \times
  \left\{ 1 + (k^2-1) \ln {\cal F}(\theta_1)
  + (k^2-1) \ln {\cal F} (\theta_2)
  + O( (k^2-1)^2 ) \right\}
\nonumber \\
  && + \frac{1}{2}
  \int_{-\pi/2+\epsilon/2}^{\pi/2-3\epsilon/2} d \theta_2
  \int_{\theta_2+\epsilon}^{\pi/2-\epsilon/2} d \theta_1~
  \left| 2 n \sin \frac{\theta_2 - \theta_1}{2 n}
  \right|^{-2 k^2}
\nonumber \\
  && \qquad \times
  \left\{ 1 + (k^2-1) \ln {\cal F}(\theta_1)
  + (k^2-1) \ln {\cal F} (\theta_2)
  + O( (k^2-1)^2 ) \right\}
\nonumber \\
  &=& \frac{1}{2}
  \int_{-\pi/2+3\epsilon/2}^{\pi/2-\epsilon/2} d \theta_2
  \int_{-\pi/2+\epsilon/2}^{\theta_2-\epsilon} d \theta_1~
  \left| 2 n \sin \frac{\theta_2 - \theta_1}{2 n}
  \right|^{-2 k^2}
\nonumber \\
  && \qquad \times
  \left\{ 1 + 2 (k^2-1) \ln {\cal F} (\theta_2)
  + O( (k^2-1)^2 ) \right\}
\nonumber \\
  && + \frac{1}{2}
  \int_{-\pi/2+\epsilon/2}^{\pi/2-3\epsilon/2} d \theta_2
  \int_{\theta_2+\epsilon}^{\pi/2-\epsilon/2} d \theta_1~
  \left| 2 n \sin \frac{\theta_2 - \theta_1}{2 n}
  \right|^{-2 k^2}
\nonumber \\
  && \qquad \times
  \left\{ 1 + 2 (k^2-1) \ln {\cal F} (\theta_2)
  + O( (k^2-1)^2 ) \right\}.
\end{eqnarray}
Next we rewrite $K_{11}(k^2)$ (\ref{K_11})
and $K_{110}(k^2) (\ref{K_110})$
similarly:
\begin{eqnarray}
  K_{11}(k^2)
  &=& \int_{-\pi/2+\epsilon/2}^{\pi/2-\epsilon/2} d \theta_2
  \int_{-\pi}^{-\pi/2-\epsilon/2} d \theta_1~
  \left| 2 \sin \frac{\theta_2 - \theta_1}{2}
  \right|^{-2 k^2}
\nonumber \\
  && \qquad \times
  \left\{ 1 + (k^2-1) \ln {\cal F}(\theta_1+\pi)
  + (k^2-1) \ln {\cal F} (\theta_2)
  + O( (k^2-1)^2 ) \right\}
\nonumber \\
  && + \int_{-\pi/2+\epsilon/2}^{\pi/2-\epsilon/2} d \theta_2
  \int_{\pi/2+\epsilon/2}^{\pi} d \theta_1~
  \left| 2 \sin \frac{\theta_2 - \theta_1}{2}
  \right|^{-2 k^2},
\nonumber \\
  && \qquad \times
  \left\{ 1 + (k^2-1) \ln {\cal F}(\theta_1-\pi)
  + (k^2-1) \ln {\cal F} (\theta_2)
  + O( (k^2-1)^2 ) \right\}
\nonumber \\
  &=& \int_{-\pi/2+\epsilon/2}^{\pi/2-\epsilon/2} d \theta_2
  \int_{-\pi}^{-\pi/2-\epsilon/2} d \theta_1~
  \left| 2 \sin \frac{\theta_2 - \theta_1}{2}
  \right|^{-2 k^2}
\nonumber \\
  && \qquad \times
  \left\{ 1 + 2 (k^2-1) \ln {\cal F} (\theta_2)
  + O( (k^2-1)^2 ) \right\}
\nonumber \\
  && + \int_{-\pi/2+\epsilon/2}^{\pi/2-\epsilon/2} d \theta_2
  \int_{\pi/2+\epsilon/2}^{\pi} d \theta_1~
  \left| 2 \sin \frac{\theta_2 - \theta_1}{2}
  \right|^{-2 k^2}
\nonumber \\
  && \qquad \times
  \left\{ 1 + 2 (k^2-1) \ln {\cal F} (\theta_2)
  + O( (k^2-1)^2 ) \right\},
\end{eqnarray}
and
\begin{eqnarray}
  K_{110}(k^2)
  &=& \int_{-\pi/2+\epsilon/2}^{\pi/2-\epsilon/2} d \theta_2
  \int_{-3 \pi/2+\epsilon/2}^{-\pi/2-\epsilon/2} d \theta_1~
  {\cal F}(\theta_1+\pi)^{k^2-1} {\cal F} (\theta_2)^{k^2-1}
  \left| 3 \sin \frac{\theta_2 - \theta_1}{3}
  \right|^{-2 k^2}
\nonumber \\
  &=& \frac{1}{2}
  \int_{-\pi/2+\epsilon/2}^{\pi/2-\epsilon/2} d \theta_2
  \int_{-3 \pi/2+\epsilon/2}^{-\pi/2-\epsilon/2} d \theta_1~
  {\cal F}(\theta_1+\pi)^{k^2-1} {\cal F} (\theta_2)^{k^2-1}
  \left| 3 \sin \frac{\theta_2 - \theta_1}{3}
  \right|^{-2 k^2}
\nonumber \\
  && + \frac{1}{2}
  \int_{-\pi/2+\epsilon/2}^{\pi/2-\epsilon/2} d \theta_2
  \int_{\pi/2+\epsilon/2}^{3 \pi/2-\epsilon/2} d \theta_1~
  {\cal F}(\theta_1-\pi)^{k^2-1} {\cal F} (\theta_2)^{k^2-1}
  \left| 3 \sin \frac{\theta_2 - \theta_1}{3}
  \right|^{-2 k^2}
\nonumber \\
  &=& \frac{1}{2}
  \int_{-\pi/2+\epsilon/2}^{\pi/2-\epsilon/2} d \theta_2
  \int_{-3 \pi/2+\epsilon/2}^{-\pi/2-\epsilon/2} d \theta_1~
  \left| 3 \sin \frac{\theta_2 - \theta_1}{3}
  \right|^{-2 k^2}
\nonumber \\
  && \qquad \times
  \left\{ 1 + (k^2-1) \ln {\cal F}(\theta_1+\pi)
  + (k^2-1) \ln {\cal F} (\theta_2)
  + O( (k^2-1)^2 ) \right\}
\nonumber \\
  && + \frac{1}{2}
  \int_{-\pi/2+\epsilon/2}^{\pi/2-\epsilon/2} d \theta_2
  \int_{\pi/2+\epsilon/2}^{3 \pi/2-\epsilon/2} d \theta_1~
  \left| 3 \sin \frac{\theta_2 - \theta_1}{3}
  \right|^{-2 k^2}
\nonumber \\
  && \qquad \times
  \left\{ 1 + (k^2-1) \ln {\cal F}(\theta_1-\pi)
  + (k^2-1) \ln {\cal F} (\theta_2)
  + O( (k^2-1)^2 ) \right\}
\nonumber \\
  &=& \frac{1}{2}
  \int_{-\pi/2+\epsilon/2}^{\pi/2-\epsilon/2} d \theta_2
  \int_{-3 \pi/2+\epsilon/2}^{-\pi/2-\epsilon/2} d \theta_1~
  \left| 3 \sin \frac{\theta_2 - \theta_1}{3}
  \right|^{-2 k^2}
\nonumber \\
  && \qquad \times
  \left\{ 1 + 2 (k^2-1) \ln {\cal F} (\theta_2)
  + O( (k^2-1)^2 ) \right\}
\nonumber \\
  && + \frac{1}{2}
  \int_{-\pi/2+\epsilon/2}^{\pi/2-\epsilon/2} d \theta_2
  \int_{\pi/2+\epsilon/2}^{3 \pi/2-\epsilon/2} d \theta_1~
  \left| 3 \sin \frac{\theta_2 - \theta_1}{3}
  \right|^{-2 k^2}
\nonumber \\
  && \qquad \times
  \left\{ 1 + 2 (k^2-1) \ln {\cal F} (\theta_2)
  + O( (k^2-1)^2 ) \right\}.
\end{eqnarray}
It is convenient to define
\begin{eqnarray}
  K_{11/110}(n, k^2)
  &\equiv& \frac{1}{2}
  \int_{-\pi/2+\epsilon/2}^{\pi/2-\epsilon/2} d \theta_2
  \int_{-n \pi+\alpha (n)}^{-\pi/2-\epsilon/2} d \theta_1~
  \left| 2 n \sin \frac{\theta_2 - \theta_1}{2 n}
  \right|^{-2 k^2}
\nonumber \\
  && \qquad \times
  \left\{ 1 + 2 (k^2-1) \ln {\cal F} (\theta_2)
  + O( (k^2-1)^2 ) \right\}
\nonumber \\
  && + \frac{1}{2}
  \int_{-\pi/2+\epsilon/2}^{\pi/2-\epsilon/2} d \theta_2
  \int_{\pi/2+\epsilon/2}^{n \pi-\alpha (n)} d \theta_1~
  \left| 2 n \sin \frac{\theta_2 - \theta_1}{2 n}
  \right|^{-2 k^2}
\nonumber \\
  && \qquad \times
  \left\{ 1 + 2 (k^2-1) \ln {\cal F} (\theta_2)
  + O( (k^2-1)^2 ) \right\},
\end{eqnarray}
where
$\alpha (n) = (n-1) \epsilon$.
Since $K_{11/110}(n, k^2)$ is related to
$K_{11}(k^2)$ and $K_{110}(k^2)$ by
\begin{equation}
  K_{11/110}(1, k^2) = \frac{1}{2} K_{11}(k^2), \quad
  K_{11/110}(3/2, k^2) = K_{110}(k^2),
\end{equation}
$K(k^2)$ is given by
\begin{equation}
  \frac{1}{2} K(k^2)
  = K_{11/110}(1, k^2) + K_{20/200}(1, k^2)
  - K_{11/110}(3/2, k^2) - K_{20/200}(3/2, k^2).
\label{four-terms}
\end{equation}
Note that
$n$-independent terms in $K_{11/110}(n, k^2)$
and $K_{20/200}(n, k^2)$ do not contribute
to $K(k^2)/2$.

As we did in Subsection 4.2,
$K_{11/110}(n, k^2) + K_{20/200}(n, k^2)$ can be
rewritten in the following factorized form
if we could neglect the divergence and set $\epsilon=0$:
\begin{eqnarray}
  && K_{11/110}(n, k^2) \bigg|_{\epsilon=0}
  + K_{20/200}(n, k^2) \bigg|_{\epsilon=0}
\nonumber \\
  && = \frac{1}{2}
  \int_{-\pi/2}^{\pi/2} d \theta_2
  \int_{-n \pi}^{n \pi} d \theta_1~
  \left| 2 n \sin \frac{\theta_2 - \theta_1}{2 n}
  \right|^{-2 k^2}
\nonumber \\
  && \qquad \times
  \left\{ 1 + 2 (k^2-1) \ln {\cal F} (\theta_2)
  + O( (k^2-1)^2 ) \right\}
\nonumber \\
  && = \frac{1}{2}
  \int_{-n \pi}^{n \pi} d \theta~
  \left| 2 n \sin \frac{\theta}{2 n}
  \right|^{-2 k^2}
\nonumber \\
  && \qquad \times
  \int_{-\pi/2}^{\pi/2} d \theta'~
  \left\{ 1 + 2 (k^2-1) \ln {\cal F} (\theta') \right\}
  + O( (k^2-1)^2 ).
\label{K-formal}
\end{eqnarray}
Let us go back to the real case with a finite $\epsilon$
and try to bring the region of the integrals
in $K_{11/110}(n, k^2) + K_{20/200}(n, k^2)$
to a form which is close to that of (\ref{K-formal}).
One such form is given by
\begin{eqnarray}
  && \int_{-\pi/2+3\epsilon/2}^{\pi/2-\epsilon/2} d \theta_2
  \int_{-\pi/2+\epsilon/2}^{\theta_2-\epsilon} d \theta_1
  + \int_{-\pi/2+\epsilon/2}^{\pi/2-3\epsilon/2} d \theta_2
  \int_{\theta_2+\epsilon}^{\pi/2-\epsilon/2} d \theta_1
\nonumber \\
  && + \int_{-\pi/2+\epsilon/2}^{\pi/2-\epsilon/2} d \theta_2
  \int_{-n \pi+\alpha (n)}^{-\pi/2-\epsilon/2} d \theta_1
  + \int_{-\pi/2+\epsilon/2}^{\pi/2-\epsilon/2} d \theta_2
  \int_{\pi/2+\epsilon/2}^{n \pi-\alpha (n)} d \theta_1
\nonumber \\
  &=& \int_{-\pi/2+\epsilon/2}^{\pi/2-\epsilon/2} d \theta_2
  \left\{ \int_{-n \pi}^{\theta_2-\epsilon} d \theta_1
  + \int_{\theta_2+\epsilon}^{n \pi} d \theta_1 \right\}
\nonumber \\
  && - \int_{-\pi/2+\epsilon/2}^{\pi/2-\epsilon/2} d \theta_2
  \left\{ \int_{-n \pi}^{-n \pi+\alpha (n)} d \theta_1
  + \int_{n \pi-\alpha (n)}^{n \pi} d \theta_1 \right\}
\nonumber \\
  && - \int_{-\pi/2+\epsilon/2}^{\pi/2-3\epsilon/2} d \theta_2
  \int_{\pi/2-\epsilon/2}^{\pi/2+\epsilon/2} d \theta_1
  - \int_{\pi/2-3\epsilon/2}^{\pi/2-\epsilon/2} d \theta_2
  \int_{\theta_2+\epsilon}^{\pi/2+\epsilon/2} d \theta_1
\nonumber \\
  && - \int_{-\pi/2+3\epsilon/2}^{\pi/2-\epsilon/2} d \theta_2
  \int_{-\pi/2-\epsilon/2}^{-\pi/2+\epsilon/2} d \theta_1
  - \int_{-\pi/2+\epsilon/2}^{-\pi/2+3\epsilon/2} d \theta_2
  \int_{-\pi/2-\epsilon/2}^{\theta_2-\epsilon} d \theta_1.
\label{K-divided}
\end{eqnarray}
The first term on the right-hand side of (\ref{K-divided})
corresponds to the region of the integrals in (\ref{K-formal})
and the remaining regions vanish in the limit $\epsilon \to 0$.

We divide $K_{11/110}(n, k^2) + K_{20/200}(n, k^2)$
into six parts 
$K_a (n, k^2)$, $K_b (n, k^2)$,
$K_c (n, k^2)$, $K_d (n, k^2)$,
$K_e (n, k^2)$, and $K_f (n, k^2)$
according to the six terms in (\ref{K-divided}):
\begin{eqnarray}
  && K_{11/110}(n, k^2) + K_{20/200}(n, k^2)
\nonumber \\
  && = K_a (n, k^2) + K_b (n, k^2)
  + K_c (n, k^2) + K_d (n, k^2)
  + K_e (n, k^2) + K_f (n, k^2).
\nonumber \\
\label{K(n)-divided}
\end{eqnarray}
For example, $K_a (n, k^2)$ is defined by
\begin{eqnarray}
  K_a (n, k^2)
  &=& \frac{1}{2}
  \int_{-\pi/2+\epsilon/2}^{\pi/2-\epsilon/2} d \theta_2
  \int_{-n \pi}^{\theta_2-\epsilon} d \theta_1
  \left| 2 n \sin \frac{\theta_2 - \theta_1}{2 n}
  \right|^{-2 k^2}
\nonumber \\
  && \qquad \times
  \left\{ 1 + 2 (k^2-1) \ln {\cal F} (\theta_2)
  + O( (k^2-1)^2 ) \right\}
\nonumber \\
  && + \frac{1}{2}
  \int_{-\pi/2+\epsilon/2}^{\pi/2-\epsilon/2} d \theta_2
  \int_{\theta_2+\epsilon}^{n \pi} d \theta_1
  \left| 2 n \sin \frac{\theta_2 - \theta_1}{2 n}
  \right|^{-2 k^2}
\nonumber \\
  && \qquad \times
  \left\{ 1 + 2 (k^2-1) \ln {\cal F} (\theta_2)
  + O( (k^2-1)^2 ) \right\}.
\end{eqnarray}
In terms of these six terms,
$K(k^2)/2$ is given by
\begin{eqnarray}
  && K(k^2)/2
\nonumber \\
  && = \left[ K_a (n, k^2) + K_b (n, k^2)
  + K_c (n, k^2) + K_d (n, k^2)
  + K_e (n, k^2) + K_f (n, k^2) \right]_{n=1}
\nonumber \\
  && - \left[ K_a (n, k^2) + K_b (n, k^2)
  + K_c (n, k^2) + K_d (n, k^2)
  + K_e (n, k^2) + K_f (n, k^2) \right]_{n=\frac{3}{2}}.
\nonumber \\
\end{eqnarray}
Let us calculate each of the six terms
on the right-hand side of (\ref{K(n)-divided}).\\

\noindent
{\boldmath $K_a (n, k^2)$}

This is most important
and factorizes as in the case of (\ref{K-formal}):
\begin{eqnarray}
  && K_a (n, k^2)
  = \frac{1}{2}
  \int_{\epsilon}^{2 n \pi -\epsilon} d \theta~
  \left| 2 n \sin \frac{\theta}{2 n}
  \right|^{-2 k^2}
\nonumber \\
  && \qquad \qquad \quad \times
  \int_{-\pi/2+\epsilon/2}^{\pi/2-\epsilon/2} d \theta'~
  \left\{ 1 + 2 (k^2-1) \ln {\cal F} (\theta') \right\}
  + O( (k^2-1)^2 ).
\end{eqnarray}
The integral over $\theta$ reduces to
the incomplete beta function
by the change of variables:
\begin{equation}
  y = \sin^2 \frac{\theta}{2 n}.
\label{theta-to-y}
\end{equation}
The result is
\begin{eqnarray}
  \int_{\epsilon}^{2 n \pi -\epsilon} d \theta~
  \left| 2 n \sin \frac{\theta}{2 n}
  \right|^{-2 k^2}
  = (2 n)^{-2 k^2+1}
  \int_{\sin^2 \frac{\epsilon}{2n}}^1 dy~
  y^{-\frac{1}{2}-k^2} (1-y)^{-\frac{1}{2}}
\nonumber \\
  = (2 n)^{-2 k^2+1}
  \left[ B \left( \frac{1}{2}-k^2, \frac{1}{2} \right)
  - B_{\sin^2 \frac{\epsilon}{2n}}
    \left( \frac{1}{2}-k^2, \frac{1}{2} \right) \right],
\label{K-1}
\end{eqnarray}
where $B_z (p,q)$ is defined by and expressed
in terms of the hypergeometric function ${}_2 F_1$ by
\begin{equation}
  B_z (p,q) = \int_0^z dt~ t^{p-1} (1-t)^{q-1}
  = \frac{z^p}{p}~ {}_2 F_1 (p, 1-q; p+1; z)
\end{equation}
for $0 < \Re (z) <1$.
What is crucial in (\ref{K-1}) is that
its divergent part when $k^2 \simeq 1$
is independent of $n$:
\begin{equation}
  (2 n)^{-2 k^2+1}
  B_{\sin^2 \frac{\epsilon}{2n}}
  \left( \frac{1}{2}-k^2, \frac{1}{2} \right)
  = \frac{2}{1-2 k^2} \epsilon^{1-2 k^2}
  \left[ 1 + O (\epsilon^2) \right].
\label{n-independence}
\end{equation}
This is easily verified
using the expression of $B_z (p,q)$
in terms of the hypergeometric function.
Therefore, unless the integral over $\theta'$
becomes too singular in the limit $\epsilon \to 0$,
the divergent part of $K_a (n, k^2)$
cancels in $K_a (1, k^2) - K_a (3/2, k^2)$,
and we find the following finite contribution:
\begin{eqnarray}
  && K_a (1, k^2) - K_a (3/2, k^2)
  = \frac{1}{2} ( 2^{-2 k^2+1}-3^{-2 k^2+1} )
  B \left( \frac{1}{2}-k^2, \frac{1}{2} \right)
\nonumber \\
  && \qquad \times
  \int_{-\pi/2+\epsilon/2}^{\pi/2-\epsilon/2} d \theta'~
  \left\{ 1 + 2 (k^2-1) \ln {\cal F} (\theta') \right\}
  + O( (k^2-1)^2 ) + o(\epsilon),
\end{eqnarray}
where $o(\epsilon)$ denotes terms which vanish
in the limit $\epsilon \to 0$
as we defined in Subsection 4.3.
Since
\begin{equation}
  B \left( \frac{1}{2}-k^2, \frac{1}{2} \right)
  = 2 \pi (k^2-1) + O( (k^2-1)^2 ),
\end{equation}
we have
\begin{equation}
  K_a (1, k^2) - K_a (3/2, k^2)
   = \frac{\pi^2}{6} (k^2-1) + O( (k^2-1)^2 )
     + o(\epsilon).
\label{K_a-final}
\end{equation}

\noindent
{\boldmath $K_b (n, k^2)$}

Since $K_b (1, k^2)$ vanishes because of $\alpha(1)=0$,
consider $K_b (3/2, k^2)$ given by
\begin{eqnarray}
  K_b (3/2, k^2)
  &=& -\frac{1}{2}
  \int_{-\pi/2+\epsilon/2}^{\pi/2-\epsilon/2} d \theta_2
  \int_{-3\pi/2}^{-3\pi/2+\epsilon/2} d \theta_1
  \left| 3 \sin \frac{\theta_2 - \theta_1}{3}
  \right|^{-2 k^2}
\nonumber \\
  && \qquad \times
  \left\{ 1 + 2 (k^2-1) \ln {\cal F} (\theta_2)
  + O( (k^2-1)^2 ) \right\}
\nonumber \\
  && - \frac{1}{2}
  \int_{-\pi/2+\epsilon/2}^{\pi/2-\epsilon/2} d \theta_2
  \int_{3\pi/2-\epsilon/2}^{3\pi/2} d \theta_1
  \left| 3 \sin \frac{\theta_2 - \theta_1}{3}
  \right|^{-2 k^2}
\nonumber \\
  && \qquad \times
  \left\{ 1 + 2 (k^2-1) \ln {\cal F} (\theta_2)
  + O( (k^2-1)^2 ) \right\}.
\end{eqnarray}
Since
\begin{equation}
  \frac{\pi}{3}
  \le \left| \frac{\theta_2-\theta_1}{3} \right|
  < \frac{2 \pi}{3},
\end{equation}
there is no singularity coming from the propagator.
The integral over $\theta_1$ is of order $\epsilon$
so that $K_b (3/2, k^2)$ vanishes unless
a compensating factor emerges from the integral
over $\theta_2$.\\

\noindent
{\boldmath $K_c (n, k^2)$}
{\bf and} {\boldmath $K_d (n, k^2)$}

Let us first consider $K_c (n, k^2)$
which is defined by
\begin{eqnarray}
  K_c (n, k^2)
  &=& -\frac{1}{2}
  \int_{-\pi/2+\epsilon/2}^{\pi/2-3\epsilon/2} d \theta_2
  \int_{\pi/2-\epsilon/2}^{\pi/2+\epsilon/2} d \theta_1
  \left| 2 n \sin \frac{\theta_2 - \theta_1}{2 n}
  \right|^{-2 k^2}
\nonumber \\
  && \qquad \times
  \left\{ 1 + 2 (k^2-1) \ln {\cal F} (\theta_2)
  + O( (k^2-1)^2 ) \right\}.
\end{eqnarray}
The integral over $\theta_1$
can be written in terms of
the incomplete beta function
by the change of variables (\ref{theta-to-y}):
\begin{eqnarray}
  && K_c (n, k^2)
  = -\frac{1}{4} (2 n)^{-2 k^2 +1}
  \int_{-\pi/2+\epsilon/2}^{\pi/2-3\epsilon/2} d \theta_2
  \left\{ 1 + 2 (k^2-1) \ln {\cal F} (\theta_2) \right\}
\nonumber \\
  && \qquad \qquad \qquad~ \times \left[
  B_{\sin^2
  \left( \frac{\pi}{4n}+\frac{\epsilon}{4n}-\frac{\theta_2}{2n}
  \right) }
  \left( \frac{1}{2}-k^2, \frac{1}{2} \right)
  - B_{\sin^2
  \left( \frac{\pi}{4n}-\frac{\epsilon}{4n}-\frac{\theta_2}{2n}
  \right) }
  \left( \frac{1}{2}-k^2, \frac{1}{2} \right)
  \right]
\nonumber \\
  && \qquad \qquad \qquad~
  + O( (k^2-1)^2 )
\nonumber \\
  && = -\frac{1}{4} (2 n)^{-2 k^2 +1}
  \int_{-\pi/2+\epsilon/2}^{\pi/2-3\epsilon/2} d \theta_2
  \left\{ 1 + 2 (k^2-1) \ln {\cal F} (\theta_2) \right\}
\nonumber \\
  && \qquad \qquad \quad \times
  B_{\sin^2
  \left( \frac{\pi}{4n}+\frac{\epsilon}{4n}-\frac{\theta_2}{2n}
  \right) }
  \left( \frac{1}{2}-k^2, \frac{1}{2} \right)
\nonumber \\
  && \quad +\frac{1}{4} (2 n)^{-2 k^2 +1}
  \int_{-\pi/2+3\epsilon/2}^{\pi/2-\epsilon/2} d \theta_2
  \left\{ 1 + 2 (k^2-1) \ln {\cal F} (\theta_2-\epsilon) \right\}
\nonumber \\
  && \qquad \qquad \quad \times
  B_{\sin^2
  \left( \frac{\pi}{4n}+\frac{\epsilon}{4n}-\frac{\theta_2}{2n}
  \right) }
  \left( \frac{1}{2}-k^2, \frac{1}{2} \right)
  + O( (k^2-1)^2 ).
\end{eqnarray}
The calculation of $K_d (n, k^2)$,
\begin{eqnarray}
  K_d (n, k^2)
  &=& -\frac{1}{2}
  \int_{\pi/2-3\epsilon/2}^{\pi/2-\epsilon/2} d \theta_2
  \int_{\theta_2+\epsilon}^{\pi/2+\epsilon/2} d \theta_1
  \left| 2 n \sin \frac{\theta_2 - \theta_1}{2 n}
  \right|^{-2 k^2}
\nonumber \\
  && \qquad \times
  \left\{ 1 + 2 (k^2-1) \ln {\cal F} (\theta_2)
  + O( (k^2-1)^2 ) \right\},
\end{eqnarray}
is almost the same as that of $K_c (n, k^2)$.
The result is
\begin{eqnarray}
  && K_d (n, k^2)
  = -\frac{1}{4} (2 n)^{-2 k^2 +1}
  \int_{\pi/2-3\epsilon/2}^{\pi/2-\epsilon/2} d \theta_2
  \left\{ 1 + 2 (k^2-1) \ln {\cal F} (\theta_2) \right\}
\nonumber \\
  && \qquad \qquad \quad \times \left[
  B_{\sin^2
  \left( \frac{\pi}{4n}+\frac{\epsilon}{4n}-\frac{\theta_2}{2n}
  \right) }
  \left( \frac{1}{2}-k^2, \frac{1}{2} \right)
  - B_{ \sin^2 \frac{\epsilon}{2n} }
  \left( \frac{1}{2}-k^2, \frac{1}{2} \right)
  \right]
  + O( (k^2-1)^2 )
\nonumber \\
  && = -\frac{1}{4} (2 n)^{-2 k^2 +1}
  \int_{\pi/2-3\epsilon/2}^{\pi/2-\epsilon/2} d \theta_2
  \left\{ 1 + 2 (k^2-1) \ln {\cal F} (\theta_2) \right\}
  B_{\sin^2
  \left( \frac{\pi}{4n}+\frac{\epsilon}{4n}-\frac{\theta_2}{2n}
  \right) }
  \left( \frac{1}{2}-k^2, \frac{1}{2} \right)
\nonumber \\
  && \quad~ +\frac{1}{4} (2 n)^{-2 k^2 +1}
  B_{ \sin^2 \frac{\epsilon}{2n} }
  \left( \frac{1}{2}-k^2, \frac{1}{2} \right)
  \int_{\pi/2-3\epsilon/2}^{\pi/2-\epsilon/2} d \theta_2
  \left\{ 1 + 2 (k^2-1) \ln {\cal F} (\theta_2) \right\}
\nonumber \\
  && \quad~
  + O( (k^2-1)^2 ).
\end{eqnarray}
The sum of $K_c (n, k^2)$ and $K_d (n, k^2)$ is given by
\begin{eqnarray}
  && K_c (n, k^2) + K_d (n, k^2)
\nonumber \\
  && = -\frac{1}{4} (2 n)^{-2 k^2 +1}
  \int_{\pi/2-3\epsilon/2}^{\pi/2-\epsilon/2} d \theta_2
  \left\{ 1 + 2 (k^2-1) \ln {\cal F} (-\theta_2) \right\}
\nonumber \\
  && \qquad \qquad \quad \times
  B_{\sin^2
  \left( \frac{\theta_2}{2n}+\frac{\pi}{4n}+\frac{\epsilon}{4n}
  \right) }
  \left( \frac{1}{2}-k^2, \frac{1}{2} \right)
\nonumber \\
  && \quad -\frac{1}{4} (2 n)^{-2 k^2 +1}
  \int_{-\pi/2+\epsilon/2}^{\pi/2-3\epsilon/2} d \theta_2
  \left\{ 2 (k^2-1)
  \ln \frac{{\cal F} (-\theta_2)}{{\cal F} (-\theta_2-\epsilon)}
  \right\}
\nonumber \\
  && \qquad \qquad \quad \times
  B_{\sin^2
  \left( \frac{\theta_2}{2n}+\frac{\pi}{4n}+\frac{\epsilon}{4n}
  \right) }
  \left( \frac{1}{2}-k^2, \frac{1}{2} \right)
\nonumber \\
  && \quad +\frac{1}{4} (2 n)^{-2 k^2 +1}
  B_{ \sin^2 \frac{\epsilon}{2n} }
  \left( \frac{1}{2}-k^2, \frac{1}{2} \right)
\nonumber \\
  && \qquad \qquad \quad \times
  \int_{\pi/2-3\epsilon/2}^{\pi/2-\epsilon/2} d \theta_2
  \left\{ 1 + 2 (k^2-1) \ln {\cal F} (\theta_2) \right\}
  + O( (k^2-1)^2 ),
\end{eqnarray}
where we redefined $-\theta_2$ as $\theta_2$
in the first two terms on the right-hand side
for later convenience.\\

\noindent
{\boldmath $K_e (n, k^2)$}
{\bf and} {\boldmath $K_f (n, k^2)$}

The calculations of $K_e (n, k^2)$ and $K_f (n, k^2)$
defined by
\begin{eqnarray}
  K_e (n, k^2)
  &=& -\frac{1}{2}
  \int_{-\pi/2+3\epsilon/2}^{\pi/2-\epsilon/2} d \theta_2
  \int_{-\pi/2-\epsilon/2}^{-\pi/2+\epsilon/2} d \theta_1
  \left| 2 n \sin \frac{\theta_2 - \theta_1}{2 n}
  \right|^{-2 k^2}
\nonumber \\
  && \qquad \times
  \left\{ 1 + 2 (k^2-1) \ln {\cal F} (\theta_2)
  + O( (k^2-1)^2 ) \right\},
\end{eqnarray}
and
\begin{eqnarray}
  K_f (n, k^2)
  &=& -\frac{1}{2}
  \int_{-\pi/2+\epsilon/2}^{-\pi/2+3\epsilon/2} d \theta_2
  \int_{-\pi/2-\epsilon/2}^{\theta_2-\epsilon} d \theta_1
  \left| 2 n \sin \frac{\theta_2 - \theta_1}{2 n}
  \right|^{-2 k^2}
\nonumber \\
  && \qquad \times
  \left\{ 1 + 2 (k^2-1) \ln {\cal F} (\theta_2)
  + O( (k^2-1)^2 ) \right\},
\end{eqnarray}
respectively, are completely parallel to
those of $K_c (n, k^2)$ and $K_d (n, k^2)$.
The sum of $K_e (n, k^2)$ and $K_f (n, k^2)$ is given by
\begin{eqnarray}
  && K_e (n, k^2) + K_f (n, k^2)
\nonumber \\
  && = -\frac{1}{4} (2 n)^{-2 k^2 +1}
  \int_{\pi/2-3\epsilon/2}^{\pi/2-\epsilon/2} d \theta_2
  \left\{ 1 + 2 (k^2-1) \ln {\cal F} (\theta_2) \right\}
\nonumber \\
  && \qquad \qquad \quad \times
  B_{\sin^2
  \left( \frac{\theta_2}{2n}+\frac{\pi}{4n}+\frac{\epsilon}{4n}
  \right) }
  \left( \frac{1}{2}-k^2, \frac{1}{2} \right)
\nonumber \\
  && \quad -\frac{1}{4} (2 n)^{-2 k^2 +1}
  \int_{-\pi/2+\epsilon/2}^{\pi/2-3\epsilon/2} d \theta_2
  \left\{ 2 (k^2-1)
  \ln \frac{{\cal F} (\theta_2)}{{\cal F} (\theta_2+\epsilon)}
  \right\}
\nonumber \\
  && \qquad \qquad \quad \times
  B_{\sin^2
  \left( \frac{\theta_2}{2n}+\frac{\pi}{4n}+\frac{\epsilon}{4n}
  \right) }
  \left( \frac{1}{2}-k^2, \frac{1}{2} \right)
\nonumber \\
  && \quad +\frac{1}{4} (2 n)^{-2 k^2 +1}
  B_{ \sin^2 \frac{\epsilon}{2n} }
  \left( \frac{1}{2}-k^2, \frac{1}{2} \right)
\nonumber \\
  && \qquad \qquad \quad \times
  \int_{-\pi/2+\epsilon/2}^{-\pi/2+3\epsilon/2} d \theta_2
  \left\{ 1 + 2 (k^2-1) \ln {\cal F} (\theta_2) \right\}
  + O( (k^2-1)^2 ).
\end{eqnarray}

Let us summarize the results.
If $\ln {\cal F} (\theta)$ is not too singular
in the limit $\epsilon \to 0$,
$K_a (1, k^2) - K_a (3/2, k^2)$ is given by (\ref{K_a-final})
and $K_b (n, k^2)$ vanishes.
The remaining four terms are combined to give
\begin{eqnarray}
  && K_c (n, k^2) + K_d (n, k^2)
   + K_e (n, k^2) + K_f (n, k^2)
\nonumber \\
  && = -\frac{1}{2} (2 n)^{-2 k^2 +1}
  \int_{\pi/2-3\epsilon/2}^{\pi/2-\epsilon/2} d \theta_2~
  B_{\sin^2
  \left( \frac{\theta_2}{2n}+\frac{\pi}{4n}+\frac{\epsilon}{4n}
  \right) }
  \left( \frac{1}{2}-k^2, \frac{1}{2} \right)
\nonumber \\
  && \qquad \qquad \quad \times
  \left\{ 1 + (k^2-1)
  \ln \left[ {\cal F} (\theta_2) {\cal F} (-\theta_2) \right]
  \right\}
\nonumber \\
  && \quad -\frac{1}{2} (2 n)^{-2 k^2 +1}
  \int_{-\pi/2+\epsilon/2}^{\pi/2-3\epsilon/2} d \theta_2~
  B_{\sin^2
  \left( \frac{\theta_2}{2n}+\frac{\pi}{4n}+\frac{\epsilon}{4n}
  \right) }
  \left( \frac{1}{2}-k^2, \frac{1}{2} \right)
\nonumber \\
  && \qquad \qquad \quad \times
  (k^2-1)
  \ln \left[
  \frac{{\cal F} (\theta_2)}{{\cal F} (\theta_2+\epsilon)}
  \frac{{\cal F} (-\theta_2)}{{\cal F} (-\theta_2-\epsilon)}
  \right]
\nonumber \\
  && \quad +\frac{1}{4} (2 n)^{-2 k^2 +1}
  B_{ \sin^2 \frac{\epsilon}{2n} }
  \left( \frac{1}{2}-k^2, \frac{1}{2} \right)
\nonumber \\
  && \qquad \qquad \quad \times
  \int_{-\pi/2+\epsilon/2}^{-\pi/2+3\epsilon/2} d \theta_2
  \left\{ 1 + 2 (k^2-1) \ln {\cal F} (\theta_2) \right\}
\nonumber \\
  && \quad +\frac{1}{4} (2 n)^{-2 k^2 +1}
  B_{ \sin^2 \frac{\epsilon}{2n} }
  \left( \frac{1}{2}-k^2, \frac{1}{2} \right)
\nonumber \\
  && \qquad \qquad \quad \times
  \int_{\pi/2-3\epsilon/2}^{\pi/2-\epsilon/2} d \theta_2
  \left\{ 1 + 2 (k^2-1) \ln {\cal F} (\theta_2)
  \right\} + O( (k^2-1)^2 ).
\label{K_c-to-K_f}
\end{eqnarray}
The incomplete beta function
in the first term on the right-hand side
does not become singular as $\theta_2 \to \pi/2$
so that this term vanishes
in the limit $\epsilon \to 0$
unless $\ln {\cal F} (\theta_2)$
becomes too singular as $\theta_2 \to \pm \pi/2$.
If $\ln {\cal F} (\theta_2)$
becomes too singular as $\theta_2 \to \pm \pi/2$,
however, point-splitting regularization we are using
will not be appropriate in the first place.

The second term
on the right-hand side of (\ref{K_c-to-K_f})
also vanishes in the limit $\epsilon \to 0$
because
\begin{equation}
  \ln \left[
  \frac{{\cal F} (\theta_2)}{{\cal F} (\theta_2+\epsilon)}
  \frac{{\cal F} (-\theta_2)}{{\cal F} (-\theta_2-\epsilon)}
  \right]
  = \left\{
  \frac{{\cal F}' (-\theta_2)}{{\cal F} (-\theta_2)}
  - \frac{{\cal F}' (\theta_2)}{{\cal F} (\theta_2)}
  \right\} \epsilon + O(\epsilon^2).
\label{ln-F}
\end{equation}
The incomplete beta function becomes singular
as $\theta_2 \to -\pi/2$, but it is not sufficient
to compensate the factor (\ref{ln-F})
unless $\ln {\cal F} (\theta_2)$
becomes too singular as $\theta_2 \to \pm \pi/2$.

As for the third and fourth terms
on the right-hand side of (\ref{K_c-to-K_f}),
their leading terms
in the limit $\epsilon \to 0$
are independent of $n$ because of (\ref{n-independence})
so that they do not contribute to $K(k^2)/2$.
It would be more difficult for the integrals over $\theta_2$
to provide compensating singular contributions than
the case of $K_a (n, k^2)$.

Therefore, only $K_a (1, k^2) - K_a (3/2, k^2)$
contributes to $K(k^2)/2$
in the limit $\epsilon \to 0$,
and $K(k^2)$ is given by
\begin{eqnarray}
  K(k^2)
  &=& 2 K_a (1, k^2) - 2 K_a (3/2, k^2) + o(\epsilon)
\nonumber \\
  &=& \frac{\pi^2}{3} (k^2-1) + O( (k^2-1)^2 )
     + o(\epsilon),
\end{eqnarray}
when $\ln {\cal F} (\theta)$ is not too singular.

Since the $\epsilon$-dependence in ${\cal F} (\theta)$
depends on the choice of the Riemann surface $\Sigma_n$
where the off-shell tachyon is defined,
it would be difficult to evaluate possible singularity
of $\ln {\cal F} (\theta)$ in general.
We can easily verify, however, that the $\ln \epsilon$
singularity in $\ln {\cal F} (\theta)$
coming from the conformal transformation $z^{1/(n-1)}$,
\begin{equation}
  \ln {\cal F} (\theta)
  = \ln \epsilon
  + \ln \frac{{\cal F}_0 (\epsilon_0 \theta/\epsilon)}
             {\epsilon_0},
\end{equation}
which follows from (\ref{F_0-to-F}),
does not change the result.

\section{Another derivation
of (\ref{vanishing-kinetic-term})}
\setcounter{equation}{0}

We found in Subsection 4.1 that
the kinetic terms
of the open string fields $\{ \varphi_i \}$
vanish when the fields satisfy
the physical state conditions
by explicit calculations.
In this appendix, we show this
using the conformal property
of physical vertex operators.

For any pair of ${\cal O}$ and ${\cal O}'$
which are primary with conformal dimension one,
let us define $\widetilde{K}$ by
\begin{equation}
  \frac{1}{2} \widetilde{K}
  = \frac{1}{2} \widetilde{K}_{11} + \widetilde{K}_{20}
    - \widetilde{K}_{110} - \widetilde{K}_{200},
\end{equation}
where
\begin{eqnarray}
  \frac{1}{2} \widetilde{K}_{11}
  &=& \frac{1}{2}
  \int_{-\pi/2+\epsilon/2}^{\pi/2-\epsilon/2} d \theta_2
  \int_{-\pi}^{-\pi/2-\epsilon/2} d \theta_1~
  \vev{ {\cal O} (e^{i \theta_1})
        {\cal O}' (e^{i \theta_2}) }_{\rm disk}
\nonumber \\
  && + \frac{1}{2}
  \int_{-\pi/2+\epsilon/2}^{\pi/2-\epsilon/2} d \theta_2
  \int_{\pi/2+\epsilon/2}^{\pi} d \theta_1~
  \vev{ {\cal O} (e^{i \theta_1})
        {\cal O}' (e^{i \theta_2}) }_{\rm disk},
\\
  \widetilde{K}_{20}
  &=& \frac{1}{2}
  \int_{-\pi/2+3\epsilon/2}^{\pi/2-\epsilon/2} d \theta_2
  \int_{-\pi/2+\epsilon/2}^{\theta_2-\epsilon} d \theta_1~
  \vev{ {\cal O} (e^{i \theta_1})
        {\cal O}' (e^{i \theta_2}) }_{\rm disk}
\nonumber \\
  && + \frac{1}{2}
  \int_{-\pi/2+\epsilon/2}^{\pi/2-3\epsilon/2} d \theta_2
  \int_{\theta_2+\epsilon}^{\pi/2-\epsilon/2} d \theta_1~
  \vev{ {\cal O} (e^{i \theta_1})
        {\cal O}' (e^{i \theta_2}) }_{\rm disk},
\\
  \widetilde{K}_{110}
  &=& \frac{1}{2}
  \int_{-\pi/2+\epsilon/2}^{\pi/2-\epsilon/2} d \theta_2
  \int_{-3 \pi/2+\epsilon/2}^{-\pi/2-\epsilon/2} d \theta_1~
  \vev{ {\cal O} (e^{i \theta_1})
        {\cal O}' (e^{i \theta_2}) }_{3 \pi}
\nonumber \\
  && + \frac{1}{2}
  \int_{-\pi/2+\epsilon/2}^{\pi/2-\epsilon/2} d \theta_2
  \int_{\pi/2+\epsilon/2}^{3\pi/2-\epsilon/2} d \theta_1~
  \vev{ {\cal O} (e^{i \theta_1})
        {\cal O}' (e^{i \theta_2}) }_{3 \pi},
\\
  \widetilde{K}_{200}
  &=& \frac{1}{2}
  \int_{-\pi/2+3\epsilon/2}^{\pi/2-\epsilon/2} d \theta_2
  \int_{-\pi/2+\epsilon/2}^{\theta_2-\epsilon} d \theta_1~
  \vev{ {\cal O} (e^{i \theta_1})
        {\cal O}' (e^{i \theta_2}) }_{3 \pi}
\nonumber \\
  && + \frac{1}{2}
  \int_{-\pi/2+\epsilon/2}^{\pi/2-3\epsilon/2} d \theta_2
  \int_{\theta_2+\epsilon}^{\pi/2-\epsilon/2} d \theta_1~
  \vev{ {\cal O} (e^{i \theta_1})
        {\cal O}' (e^{i \theta_2}) }_{3 \pi}.
\end{eqnarray}
These definitions are obvious generalizations of $K(1)$,
$K_{11}(1)$, $K_{20}(1)$, $K_{110}(1)$, and $K_{200}(1)$
we studied in Subsection 4.1 and Appendix B.
Using (\ref{K-divided}) and the argument
in Appendix B showing that $K_b (n, k^2)$,
$K_c (n, k^2)$, $K_d (n, k^2)$,
$K_e (n, k^2)$, and $K_f (n, k^2)$
vanish in the limit $\epsilon \to 0$,
we can show that $\widetilde{K}_{11}/2+\widetilde{K}_{20}$
can be written as
\begin{eqnarray}
  \frac{1}{2} \widetilde{K}_{11} + \widetilde{K}_{20}
  &=& \frac{1}{2}
  \int_{-\pi/2+\epsilon/2}^{\pi/2-\epsilon/2} d \theta_2
  \int_{-\pi}^{\theta_2-\epsilon} d \theta_1
  \vev{ {\cal O} (e^{i \theta_1})
        {\cal O}' (e^{i \theta_2}) }_{\rm disk}
\nonumber \\
  && + \frac{1}{2}
  \int_{-\pi/2+\epsilon/2}^{\pi/2-\epsilon/2} d \theta_2
  \int_{\theta_2+\epsilon}^{\pi} d \theta_1
  \vev{ {\cal O} (e^{i \theta_1})
        {\cal O}' (e^{i \theta_2}) }_{\rm disk}
  + o(\epsilon)
\nonumber \\
  &=& \frac{1}{2}
  \int_{-\pi/2+\epsilon/2}^{\pi/2-\epsilon/2} d \theta_2
  \int_{\theta_2+\epsilon}^{\theta_2+2\pi-\epsilon}
  d \theta_1
  \vev{ {\cal O} (e^{i \theta_1})
        {\cal O}' (e^{i \theta_2}) }_{\rm disk}
  + o(\epsilon)
\nonumber \\
  &=& \frac{\pi-\epsilon}{2}
  \int_{\epsilon}^{2\pi-\epsilon} d \theta
  \vev{ {\cal O} (e^{i \theta}) {\cal O}' (1) }_{\rm disk}
  + o(\epsilon),
\end{eqnarray}
where we used
\begin{equation}
  \vev{ {\cal O} (e^{i \theta_1})
        {\cal O}' (e^{i \theta_2}) }_{\rm disk}
  =  \vev{ {\cal O} (e^{i (\theta_1-\theta_2)})
         {\cal O}' (1) }_{\rm disk}.
\end{equation}
Similarly for $\widetilde{K}_{110}+\widetilde{K}_{200}$,
we have
\begin{eqnarray}
  \widetilde{K}_{110}+\widetilde{K}_{200}
  &=& \frac{1}{2}
  \int_{-\pi/2+\epsilon/2}^{\pi/2-\epsilon/2} d \theta_2
  \int_{-3\pi/2}^{\theta_2-\epsilon} d \theta_1
  \vev{ {\cal O} (e^{i \theta_1})
        {\cal O}' (e^{i \theta_2}) }_{3 \pi}
\nonumber \\
  && + \frac{1}{2}
  \int_{-\pi/2+\epsilon/2}^{\pi/2-\epsilon/2} d \theta_2
  \int_{\theta_2+\epsilon}^{3\pi/2} d \theta_1
  \vev{ {\cal O} (e^{i \theta_1})
        {\cal O}' (e^{i \theta_2}) }_{3 \pi}
  + o(\epsilon).
\end{eqnarray}
Let us make the conformal transformation $z^{2/3}$
for $\widetilde{K}_{110}+\widetilde{K}_{200}$.
Since ${\cal O}$ and ${\cal O}'$ are conformal primary
with dimension one,
$\widetilde{K}_{110}+\widetilde{K}_{200}$ is transformed as
\begin{eqnarray}
  \widetilde{K}_{110}+\widetilde{K}_{200}
  &=& \frac{1}{2}
  \int_{-\pi/3+\epsilon/3}^{\pi/3-\epsilon/3} d \theta_2
  \int_{-\pi}^{\theta_2-2\epsilon/3} d \theta_1
  \vev{ {\cal O} (e^{i \theta_1})
        {\cal O}' (e^{i \theta_2}) }_{\rm disk}
\nonumber \\
  && + \frac{1}{2}
  \int_{-\pi/3+\epsilon/3}^{\pi/3-\epsilon/3} d \theta_2
  \int_{\theta_2+2\epsilon/3}^{\pi} d \theta_1
  \vev{ {\cal O} (e^{i \theta_1})
        {\cal O}' (e^{i \theta_2}) }_{\rm disk}
  + o(\epsilon)
\nonumber \\
  &=& \frac{1}{2}
  \int_{-\pi/3+\epsilon/3}^{\pi/3-\epsilon/3} d \theta_2
  \int_{\theta_2+2\epsilon/3}^{\theta_2+2\pi-2\epsilon/3}
  d \theta_1
  \vev{ {\cal O} (e^{i \theta_1})
        {\cal O}' (e^{i \theta_2}) }_{\rm disk}
  + o(\epsilon)
\nonumber \\
  &=& \frac{\pi-\epsilon}{3}
  \int_{2\epsilon/3}^{2\pi-2\epsilon/3} d \theta
  \vev{ {\cal O} (e^{i \theta}) {\cal O}' (1) }_{\rm disk}
  + o(\epsilon).
\label{K_110+K_200-transformed}
\end{eqnarray}
Let us make a further conformal transformation
to (\ref{K_110+K_200-transformed})
such that the integral is from $\epsilon$
to $2\pi-\epsilon$.
Consider a class of conformal transformations
parametrized by a real, positive constant $a$,
\begin{equation}
  f(z) = \frac{(1+a)z+1-a}{(1-a)z+1+a}, \quad
  \left. \frac{df(z)}{dz} \right|_{z=1} = a,
\end{equation}
which maps the unit disk to itself:
\begin{equation}
  | f(e^{i \theta}) | = 1, \quad
  | f(0) | < 1.
\end{equation}
The constant $a$ is determined as
\begin{equation}
  a = \frac{\tan \frac{\epsilon}{2}}
           {\tan \frac{\epsilon}{3}}
\end{equation}
by the condition that
\begin{equation}
  f(e^{\pm 2i\epsilon/3}) = e^{\pm i \epsilon}.
\end{equation}
While the integral of ${\cal O} (e^{i \theta})$
over $\theta$ remains invariant,
the operator ${\cal O}' (1)$ is transformed as
\begin{equation}
  {\cal O}' (1) \to
  \frac{\tan \frac{\epsilon}{2}}
       {\tan \frac{\epsilon}{3}}~ {\cal O}' (1).
\end{equation}
By this conformal transformation, the last line of
(\ref{K_110+K_200-transformed}) is transformed as
\begin{equation}
  \widetilde{K}_{110} + \widetilde{K}_{200}
  = \frac{\pi-\epsilon}{3}~
               \frac{\tan \frac{\epsilon}{2}}
                    {\tan \frac{\epsilon}{3}}
  \int_{\epsilon}^{2\pi-\epsilon} d \theta
  \vev{ {\cal O} (e^{i \theta}) {\cal O}' (1) }_{\rm disk}
  + o(\epsilon).
\end{equation}
Therefore, $\widetilde{K}/2$ is given by
\begin{equation}
  \frac{1}{2} \widetilde{K}
  = \frac{\pi-\epsilon}{2}
    \left( 1 - \frac{2}{3}
               \frac{\tan \frac{\epsilon}{2}}
                    {\tan \frac{\epsilon}{3}} \right)
  \int_{\epsilon}^{2\pi-\epsilon} d \theta
  \vev{ {\cal O} (e^{i \theta}) {\cal O}' (1) }_{\rm disk}
  + o(\epsilon).
\end{equation}
Since ${\cal O}$ and ${\cal O}'$ are primary
with conformal dimension one,
the singularity of the propagator is
\begin{equation}
  \vev{ {\cal O} (e^{i \theta_1})
        {\cal O}' (e^{i \theta_2}) }_{\rm disk}
  \sim \frac{1}{(\theta_1-\theta_2)^2}
\end{equation}
when $\theta_1 \sim \theta_2$ so that
\begin{equation}
  \int_{\epsilon}^{2\pi-\epsilon} d \theta
  \vev{ {\cal O} (e^{i \theta}) {\cal O}' (1) }_{\rm disk}
  = O \left( \frac{1}{\epsilon} \right).
\end{equation}
On the other hand,
the factor in front of the integral scales
in the limit $\epsilon \to 0$ as
\begin{equation}
  1 - \frac{2}{3}
      \frac{\tan \frac{\epsilon}{2}}
           {\tan \frac{\epsilon}{3}}
  = O(\epsilon^2),
\end{equation}
therefore
\begin{equation}
  \frac{1}{2} \widetilde{K} = o(\epsilon).
\end{equation}
This explains that the cancellation of the finite terms
between (\ref{finite-term-1}) and (\ref{finite-term-2})
in Subsection 4.1 is not accidental
but a consequence of the conformal property
of the physical vertex operators.

\section{$V_{30} + V_{21} - V_{300} - V_{210} - V_{201}$}
\setcounter{equation}{0}

We calculate $V_{30}$, $V_{300}$,
$V_{21}$, $V_{210}$, and $V_{201}$
defined in Subsection 4.3
to show that
$V_{30} + V_{21} - V_{300} - V_{210} - V_{201}$
vanishes in the limit $\epsilon \to 0$.\\

\noindent
{\bf Definitions}

We regularize the state $\ket{\it 3}$
by regularizing the integrals of the inserted vertex operators
in the cone representation as follows:
\begin{eqnarray}
  && \int_{-(n-1) \pi/2 +5 \epsilon_0/2}^{(n-1) \pi/2
  -\epsilon_0/2} d \theta_3
  \int_{-(n-1) \pi/2 +3 \epsilon_0/2}^{\theta_3-\epsilon_0}
  d \theta_2
  \int_{-(n-1) \pi/2 +\epsilon_0/2}^{\theta_2-\epsilon_0}
  d \theta_1
\nonumber \\
  && \qquad \times
  {\cal F}_0 (\theta_1)^{k^2-1} e^{ikX (e^{i \theta_1})}
  {\cal F}_0 (\theta_2)^{k^2-1} e^{ikX (e^{i \theta_2})}
  {\cal F}_0 (\theta_3)^{k^2-1} e^{ikX (e^{i \theta_3})}.
\end{eqnarray}
The propagators in $V_{30}$ and $V_{300}$ are given
in a similar way as
in the case of $V_{111}$ in (\ref{V_111}).
It is convenient to define $V_{30/300}(n)$ by
\begin{eqnarray}
  && V_{30/300}(n)
  \equiv \int_{5\epsilon/2}^{\pi-\epsilon/2} d \theta_3
  \int_{3\epsilon/2}^{\theta_3-\epsilon} d \theta_2
  \int_{\epsilon/2}^{\theta_2-\epsilon} d \theta_1
\nonumber \\
  && \qquad \qquad \times
  \left( 2 n \sin \frac{\theta_2-\theta_1}{2 n} \right)^{-1}
  \left( 2 n \sin \frac{\theta_3-\theta_1}{2 n} \right)^{-1}
  \left( 2 n \sin \frac{\theta_3-\theta_2}{2 n} \right)^{-1}.
\end{eqnarray}
This is related to $V_{30}$ and $V_{300}$ as
\begin{equation}
  V_{30} = V_{30/300}(1), \quad V_{300} = V_{30/300}(3/2).
\end{equation}
As for $V_{21}$, $V_{210}$, and $V_{201}$,
we define $V_{21/210+201}(n)$ by
\begin{eqnarray}
  && V_{21/210+201}(n)
  \equiv \int_{C_n} d \theta_3
  \int_{-\pi+3\epsilon/2}^{-\epsilon/2} d \theta_2
  \int_{-\pi+\epsilon/2}^{\theta_2-\epsilon} d \theta_1
\nonumber \\
  && \qquad \qquad \times
  \left( 2 n \sin \frac{\theta_2-\theta_1}{2 n} \right)^{-1}
  \left( 2 n \sin \frac{\theta_3-\theta_1}{2 n} \right)^{-1}
  \left( 2 n \sin \frac{\theta_3-\theta_2}{2 n} \right)^{-1},
\end{eqnarray}
where
\begin{equation}
  \int_{C_1} d \theta_3
  = \int_{\epsilon/2}^{\pi-\epsilon/2}, \qquad
  \int_{C_{3/2}} d \theta_3
  = \int_{\epsilon/2}^{\pi-\epsilon/2}
  + \int_{\pi+\epsilon/2}^{2\pi-\epsilon/2}.
\end{equation}
This is related to $V_{21}$ and $V_{210}+V_{201}$ as
\begin{equation}
  V_{21} = V_{21/210+201}(1), \quad
  V_{210}+V_{201} = V_{21/210+201}(3/2).
\end{equation}
The contour $C_n$ can also be written as
\begin{equation}
  \int_{C_n} d \theta_3
  = \int_{\epsilon/2}^{(2n-1)\pi-\epsilon/2} d \theta_3
  +  \int_{\widetilde{C}_n} d \theta_3,
\label{C-divided}
\end{equation}
where
\begin{equation}
  \int_{\widetilde{C}_1} d \theta_3 = 0, \qquad
  \int_{\widetilde{C}_{3/2}} d \theta_3
  = \int_{\pi+\epsilon/2}^{\pi-\epsilon/2} d \theta_3.
\label{C-tilde}
\end{equation}
As we will see, integrals along $\widetilde{C}_n$
are unimportant in most cases.

Using $V_{30/300}(n)$ and $V_{21/210+201}(n)$,
$V_{30} + V_{21} - V_{300} - V_{210} - V_{201}$
is expressed as
\begin{eqnarray}
  && V_{30} + V_{21} - V_{300} - V_{210} - V_{201}
\nonumber \\
  &=& V_{30/300}(1) + V_{21/210+201}(1)
  - V_{30/300}(3/2) - V_{21/210+201}(3/2).
\label{V-to-V(n)}
\end{eqnarray}
Let us calculate $V_{30/300}(n)$
and $V_{21/210+201}(n)$.\\

\noindent
{\boldmath $V_{30/300}(n)$}

The calculations of
the integrals over $\theta_1$ and $\theta_2$ are
tedious but straightforward:
\begin{eqnarray}
  && V_{30/300}(n)
\nonumber \\
  && = \frac{1}{(2n)^2}
  \int_{5\epsilon/2}^{\pi-\epsilon/2} d \theta_3
  \int_{3\epsilon/2}^{\theta_3-\epsilon} d \theta_2
  \left( \sin \frac{\theta_3-\theta_2}{2 n} \right)^{-2}
  \ln \left(
  \frac{\sin \frac{\theta_3-\theta_2+\epsilon}{2 n}}
       {\sin \frac{\epsilon}{2 n}}
  \frac{\sin \frac{\theta_2-\epsilon/2}{2 n}}
       {\sin \frac{\theta_3-\epsilon/2}{2 n}}
  \right)
\nonumber \\
  && = \frac{1}{n}
  \int_{5\epsilon/2}^{\pi-\epsilon/2} d \theta_3
  \Bigg\{
  - \frac{\theta_3-5\epsilon/2}{2n}
  + \cot \frac{\theta_3-\epsilon/2}{2n}~
    \ln \frac{\sin \frac{\epsilon}{2 n}}
             {\sin \frac{\theta_3-3\epsilon/2}{2 n}}
\nonumber \\
  && \qquad \qquad \qquad \qquad \qquad \qquad
  + \cot \frac{\epsilon}{2 n}~
  \ln \left(
  \frac{\sin \frac{\epsilon}{n}}
       {\sin \frac{\epsilon}{2 n}}
  \frac{\sin \frac{\theta_3-3\epsilon/2}{2 n}}
       {\sin \frac{\theta_3-\epsilon/2}{2 n}}
  \right)
  \Bigg\}.
\end{eqnarray}
We divide this expression into the following six terms:
\begin{eqnarray}
  && V_{30/300}(n)
  = -\frac{1}{n}
  \int_{5\epsilon/2}^{\pi-\epsilon/2} d \theta_3~
  \frac{\theta_3-5\epsilon/2}{2n}
  + \frac{1}{n}
  \int_{5\epsilon/2}^{\pi-\epsilon/2} d \theta_3~
  \cot \frac{\theta_3-\epsilon/2}{2n}~
  \ln \sin \frac{\epsilon}{2 n}
\nonumber \\
  && \quad - \frac{1}{n}
  \int_{5\epsilon/2}^{\pi-\epsilon/2} d \theta_3~
  \cot \frac{\theta_3-\epsilon/2}{2n}~
  \ln \frac{\sin \frac{\theta_3-3\epsilon/2}{2 n}}
           {\sin \frac{\theta_3-\epsilon/2}{2 n}}
\nonumber \\
  && \quad - \frac{1}{n}
  \int_{5\epsilon/2}^{\pi-\epsilon/2} d \theta_3~
  \cot \frac{\theta_3-\epsilon/2}{2n}~
  \ln \sin \frac{\theta_3-\epsilon/2}{2 n}
\nonumber \\
  && \quad + \frac{1}{n}
  \int_{5\epsilon/2}^{\pi-\epsilon/2} d \theta_3~
  \cot \frac{\epsilon}{2 n}~
    \ln \frac{\sin \frac{\epsilon}{n}}
             {\sin \frac{\epsilon}{2 n}}
  + \frac{1}{n}
  \int_{5\epsilon/2}^{\pi-\epsilon/2} d \theta_3~
  \cot \frac{\epsilon}{2n}~
  \ln \frac{\sin \frac{\theta_3-3\epsilon/2}{2 n}}
           {\sin \frac{\theta_3-\epsilon/2}{2 n}}.
\label{V_30/300-6}
\end{eqnarray}
The first, second, fourth, and fifth integrals
are easily carried out to give
\begin{eqnarray}
  && -\frac{1}{n}
  \int_{5\epsilon/2}^{\pi-\epsilon/2} d \theta_3~
  \frac{\theta_3-5\epsilon/2}{2n}
  = -\frac{\pi^2}{(2n)^2} + o(\epsilon),
\nonumber \\
  && \frac{1}{n}
  \int_{5\epsilon/2}^{\pi-\epsilon/2} d \theta_3~
  \cot \frac{\theta_3-\epsilon/2}{2n}~
  \ln \sin \frac{\epsilon}{2 n}
  = 2 \ln \sin \frac{\epsilon}{2 n}~
  \ln \frac{\sin \frac{\pi}{2 n}}
           {\sin \frac{\epsilon}{n}} + o(\epsilon),
\nonumber \\
  && - \frac{1}{n}
  \int_{5\epsilon/2}^{\pi-\epsilon/2} d \theta_3~
  \cot \frac{\theta_3-\epsilon/2}{2n}~
  \ln \sin \frac{\theta_3-\epsilon/2}{2 n}
  = \left( \ln \sin \frac{\epsilon}{n} \right)^2
  - \left( \ln \sin \frac{\pi}{2 n} \right)^2
  + o(\epsilon),
\nonumber \\
  && \frac{1}{n}
  \int_{5\epsilon/2}^{\pi-\epsilon/2} d \theta_3~
  \cot \frac{\epsilon}{2 n}~
    \ln \frac{\sin \frac{\epsilon}{n}}
             {\sin \frac{\epsilon}{2 n}}
  = \frac{2 \pi \ln 2}{\epsilon} - 6 \ln 2
    + o(\epsilon).
\end{eqnarray}
The third integral in (\ref{V_30/300-6}) is
finite and independent of $n$
in the limit $\epsilon \to 0$.
This can be seen by the following change of
variables:
\begin{eqnarray}
  - \frac{1}{n}
  \int_{5\epsilon/2}^{\pi-\epsilon/2} d \theta_3~
  \cot \frac{\theta_3-\epsilon/2}{2n}~
  \ln \frac{\sin \frac{\theta_3-3\epsilon/2}{2 n}}
           {\sin \frac{\theta_3-\epsilon/2}{2 n}}
  = -2 \int_{\frac{2 n \epsilon}{\pi - \epsilon}}^{n} dx~
  \frac{\epsilon}{x^2} \cot \frac{\epsilon}{x}~
  \ln \frac{\sin \left( 
                 \frac{\epsilon}{x} - \frac{\epsilon}{2n}
                 \right)}
           {\sin \frac{\epsilon}{x}}.
\nonumber \\
\end{eqnarray}
We can take the limit $\epsilon \to 0$ to find
\begin{eqnarray}
 &&  -2 \int_{0}^{n} \frac{dx}{x}~
  \ln \frac{ \frac{1}{x} - \frac{1}{2n}}{\frac{1}{x}}
  + o(\epsilon)
  = -2 \int_0^1 \frac{dx}{x}~
  \ln \left( 1 - \frac{x}{2} \right)
  + o(\epsilon)
\nonumber \\
  &=& 2~ {\rm Li}_2 \left( \frac{1}{2} \right) + o(\epsilon)
  = \frac{\pi^2}{6} - (\ln 2)^2 + o(\epsilon),
\label{V_30/300-3-final}
\end{eqnarray}
where the polylogarithm ${\rm Li}_2 (z)$ is defined by
\begin{equation}
  {\rm Li}_2 (z) = -\int_0^z dt~ \frac{\ln (1-t)}{t}.
\label{polylogarithm}
\end{equation}
The finite value in (\ref{V_30/300-3-final})
is not important
because it is independent of $n$.
Finally, we rewrite the last integral
in (\ref{V_30/300-6}) as follows:
\begin{eqnarray}
  && \frac{1}{n}
  \int_{5\epsilon/2}^{\pi-\epsilon/2} d \theta_3~
  \cot \frac{\epsilon}{2n}~
  \ln \frac{\sin \frac{\theta_3-3\epsilon/2}{2 n}}
           {\sin \frac{\theta_3-\epsilon/2}{2 n}}
\nonumber \\
  && = \frac{1}{n}
  \cot \frac{\epsilon}{2n}
  \left\{
  \int_{\epsilon}^{2\epsilon} d \theta~
  \ln \sin \frac{\theta}{2 n}
  - \int_{\pi-2\epsilon}^{\pi-\epsilon} d \theta~
  \ln \sin \frac{\theta}{2 n}
  \right\}
\nonumber \\
  && = \frac{1}{n}
  \cot \frac{\epsilon}{2n}
  \int_{\epsilon}^{2\epsilon} d \theta~
  \ln \sin \frac{\theta}{2 n}
  - 2 \ln \sin \frac{\pi}{2 n}
  + o(\epsilon).
\end{eqnarray}
The final result for $V_{30/300}(n)$ is given by
\begin{eqnarray}
  V_{30/300}(n)
  &=& -\frac{\pi^2}{(2n)^2}
  + 2 \ln \sin \frac{\epsilon}{2 n}~
  \ln \frac{\sin \frac{\pi}{2 n}}
           {\sin \frac{\epsilon}{n}}
\nonumber \\
  && + \frac{\pi^2}{6} - (\ln 2)^2
  + \left( \ln \sin \frac{\epsilon}{n} \right)^2
  - \left( \ln \sin \frac{\pi}{2 n} \right)^2
  + \frac{2 \pi \ln 2}{\epsilon} - 6 \ln 2
\nonumber \\
  && + \frac{1}{n}
  \cot \frac{\epsilon}{2n}
  \int_{\epsilon}^{2\epsilon} d \theta~
  \ln \sin \frac{\theta}{2 n}
  - 2 \ln \sin \frac{\pi}{2 n}
  + o(\epsilon).
\end{eqnarray}
\\

\noindent
{\boldmath $V_{21/210+201}(n)$}

Let us move on to $V_{21/210+201}(n)$.
The calculations of
the integrals over $\theta_1$ and $\theta_2$ are
again tedious but straightforward:
\begin{eqnarray}
  && V_{21/210+201}(n)
\nonumber \\
  && = \frac{1}{(2n)^2} \int_{C_n} d \theta_3
  \int_{-\pi+3\epsilon/2}^{-\epsilon/2} d \theta_2
  \left( \sin \frac{\theta_3-\theta_2}{2 n} \right)^{-2}
  \ln \left(
  \frac{\sin \frac{\theta_3-\theta_2+\epsilon}{2 n}}
       {\sin \frac{\epsilon}{2 n}}
  \frac{\sin \frac{\theta_2+\pi-\epsilon/2}{2 n}}
       {\sin \frac{\theta_3+\pi-\epsilon/2}{2 n}}
  \right)
\nonumber \\
  && = \frac{1}{n}
  \int_{C_n} d \theta_3~
  \Bigg\{
  - \frac{\pi-2\epsilon}{2n}
  + \cot \frac{\theta_3+\epsilon/2}{2 n}~
  \ln \left(
  \frac{\sin \frac{\pi-\epsilon}{2 n}}
       {\sin \frac{\epsilon}{2 n}}
  \frac{\sin \frac{\theta_3+3\epsilon/2}{2 n}}
       {\sin \frac{\theta_3+\pi-\epsilon/2}{2 n}}
  \right)
\nonumber \\
  && \qquad \qquad \qquad \qquad \qquad \qquad
  + \cot \frac{\epsilon}{2n}~
    \ln \frac{\sin \frac{\theta_3+3\epsilon/2}{2 n}}
             {\sin \frac{\theta_3+\epsilon/2}{2 n}}
  \Bigg\}.
\end{eqnarray}
We divide this expression into the following six terms:
\begin{eqnarray}
  && V_{21/210+201}(n)
  = - \frac{1}{n}
  \int_{C_n} d \theta_3~
  \frac{\pi-2\epsilon}{2n}
  + \frac{1}{n}
  \int_{C_n} d \theta_3~
  \cot \frac{\theta_3+\epsilon/2}{2 n}~
  \ln \frac{\sin \frac{\pi-\epsilon}{2 n}}
           {\sin \frac{\epsilon}{2 n}}
\nonumber \\
  && + \frac{1}{n}
  \int_{C_n} d \theta_3~
  \cot \frac{\theta_3+\epsilon/2}{2 n}~
  \ln \frac{\sin \frac{\theta_3+3\epsilon/2}{2 n}}
           {\sin \frac{\theta_3+\epsilon/2}{2 n}}
  + \frac{1}{n}
  \int_{C_n} d \theta_3~
  \cot \frac{\theta_3+\epsilon/2}{2 n}~
  \ln \sin \frac{\theta_3+\epsilon/2}{2 n}
\nonumber \\
  && - \frac{1}{n}
  \int_{C_n} d \theta_3~
  \cot \frac{\theta_3+\epsilon/2}{2 n}~
  \ln \sin \frac{\theta_3+\pi-\epsilon/2}{2 n}
\nonumber \\
  && + \frac{1}{n}
  \int_{C_n} d \theta_3~
  \cot \frac{\epsilon}{2n}~
  \ln \frac{\sin \frac{\theta_3+3\epsilon/2}{2 n}}
           {\sin \frac{\theta_3+\epsilon/2}{2 n}}.
\label{V_21/210+201-6}
\end{eqnarray}
The first, second, and fourth integrals
in (\ref{V_21/210+201-6}) are easily carried out.
The contributions from the contour $\widetilde{C}_n$
defined by (\ref{C-tilde}) to these integrals vanish
in the limit $\epsilon \to 0$
so that we can replace the contour $C_n$
by the first term
on the right-hand side of (\ref{C-divided}).
The three integrals are carried out as follows:
\begin{eqnarray}
  && - \frac{1}{n}
  \int_{\epsilon/2}^{(2n-1)\pi-\epsilon/2} d \theta_3~
  \frac{\pi-2\epsilon}{2n}
  = -\frac{(2n-1) \pi^2}{2 n^2} + o(\epsilon),
\nonumber \\
  && \frac{1}{n}
  \int_{\epsilon/2}^{(2n-1)\pi-\epsilon/2} d \theta_3~
  \cot \frac{\theta_3+\epsilon/2}{2 n}~
  \ln \frac{\sin \frac{\pi-\epsilon}{2 n}}
           {\sin \frac{\epsilon}{2 n}}
  = 2 \left( \ln \frac{\sin \frac{\pi}{2 n}}
                      {\sin \frac{\epsilon}{2 n}} \right)^2
    + o(\epsilon),
\nonumber \\
  && \frac{1}{n}
  \int_{\epsilon/2}^{(2n-1)\pi-\epsilon/2} d \theta_3~
  \cot \frac{\theta_3+\epsilon/2}{2 n}~
  \ln \sin \frac{\theta_3+\epsilon/2}{2 n}
  = \left( \ln \sin \frac{\pi}{2n} \right)^2
    - \left( \ln \sin \frac{\epsilon}{2n} \right)^2.
\nonumber \\
\end{eqnarray}
The third integral in (\ref{V_21/210+201-6}) is finite
and independent of $n$.
We can show this in a similar way as in the case of
the third integral in (\ref{V_30/300-6}).
The contribution from the contour $\widetilde{C}_n$
vanishes in the limit $\epsilon \to 0$,
and the remaining part gives
\begin{eqnarray}
  && \frac{1}{n}
  \int_{\epsilon/2}^{(2n-1)\pi-\epsilon/2} d \theta_3~
  \cot \frac{\theta_3+\epsilon/2}{2 n}~
  \ln \frac{\sin \frac{\theta_3+3\epsilon/2}{2 n}}
           {\sin \frac{\theta_3+\epsilon/2}{2 n}}
\nonumber \\
  && = 2 \int_{\frac{2n \epsilon}{(2n-1) \pi}}^{2n} dx~
    \frac{\epsilon}{x^2} \cot \frac{\epsilon}{x}~
    \ln \frac{\sin \left(
                   \frac{\epsilon}{x}+\frac{\epsilon}{2 n}
                   \right)}
             {\sin \frac{\epsilon}{x}}
\nonumber \\
  && = 2 \int_0^2 \frac{dx}{x}~
    \ln \left( 1 + \frac{x}{2} \right) + o(\epsilon)
  = 2~ {\rm Li}_2 (-1) + o(\epsilon)
  = \frac{\pi^2}{6} + o(\epsilon).
\end{eqnarray}
The $n$-independent finite value here is again unimportant.

For the fifth integral in (\ref{V_21/210+201-6}),
we could not calculate it for a generic value of $n$.
In the case of $n=1$, the integral is finite
in the limit $\epsilon \to 0$ and given by
\begin{eqnarray}
  && - \int_{\epsilon/2}^{\pi-\epsilon/2} d \theta_3~
  \cot \frac{\theta_3+\epsilon/2}{2}~
  \ln \sin \frac{\theta_3+\pi-\epsilon/2}{2}
  = - \int_{0}^{\pi} d \theta_3~
  \cot \frac{\theta_3}{2}~
  \ln \cos \frac{\theta_3}{2} + o(\epsilon)
\nonumber \\
  && = -2 \int_{0}^{1} \frac{dx}{x} \ln \sqrt{1-x^2}
    + o(\epsilon)
  = {\rm Li}_2 (1) + {\rm Li}_2 (-1) + o(\epsilon)
  = \frac{\pi^2}{12} + o(\epsilon),
\end{eqnarray}
where we changed the variable as $x = \sin (\theta/2)$.
For $n=3/2$, we have
\begin{eqnarray}
  && - \frac{2}{3}
  \int_{\epsilon/2}^{\pi-\epsilon/2} d \theta_3~
  \cot \frac{\theta_3+\epsilon/2}{3}~
  \ln \sin \frac{\theta_3+\pi-\epsilon/2}{3}
\nonumber \\
  && - \frac{2}{3}
  \int_{\pi+\epsilon/2}^{2\pi-\epsilon/2} d \theta_3~
  \cot \frac{\theta_3+\epsilon/2}{3}~
  \ln \sin \frac{\theta_3+\pi-\epsilon/2}{3}
\nonumber \\
  && = - \left[
  2 \ln \sin \frac{\theta_3+\epsilon/2}{3}~
  \ln \sin \frac{\theta_3+\pi-\epsilon/2}{3}
  \right]_{\theta_3=\epsilon/2}^{\theta_3=\pi-\epsilon/2}
\nonumber \\
  && + \frac{2}{3}
  \int_{\epsilon/2}^{\pi-\epsilon/2} d \theta_3~
  \ln \sin \frac{\theta_3+\epsilon/2}{3}~
  \cot \frac{\theta_3+\pi-\epsilon/2}{3}
\nonumber \\
  && - \frac{2}{3}
  \int_{\pi+\epsilon/2}^{2\pi-\epsilon/2} d \theta_3~
  \cot \frac{\theta_3+\epsilon/2}{3}~
  \ln \sin \frac{\theta_3+\pi-\epsilon/2}{3}
\nonumber \\
  && = -2 \ln \sin \frac{\pi-\epsilon}{3}~
  \ln \sin \frac{2\pi-\epsilon}{3}
  +2 \ln \sin \frac{\epsilon}{3}~
  \ln \sin \frac{\pi}{3}
\nonumber \\
  && + \frac{4}{3}
  \int_{\epsilon/2}^{\pi-\epsilon/2} d \theta_3~
  \ln \sin \frac{\theta_3+\epsilon/2}{3}~
  \cot \frac{\theta_3+\pi-\epsilon/2}{3}
\nonumber \\
  && = -2 \left( \ln \sin \frac{\pi}{3} \right)^2
  +2 \ln \sin \frac{\epsilon}{3}~
  \ln \sin \frac{\pi}{3}
  + \frac{4}{3}
  \int_{0}^{\pi} d \theta~
  \cot \frac{\theta+\pi}{3}
  \ln \sin \frac{\theta}{3} + o(\epsilon).
\nonumber \\
\end{eqnarray}
The calculation of the integral in the last line
is slightly complicated so that we postpone it
and finish the remaining part of $V_{21/210+201}(n)$.

For the last integral in (\ref{V_21/210+201-6}),
we divide it into two parts according to (\ref{C-divided}).
For the first contour, we have
\begin{eqnarray}
  && \frac{1}{n}
  \int_{\epsilon/2}^{(2n-1)\pi-\epsilon/2} d \theta_3~
  \cot \frac{\epsilon}{2n}~
  \ln \frac{\sin \frac{\theta_3+3\epsilon/2}{2 n}}
           {\sin \frac{\theta_3+\epsilon/2}{2 n}}
\nonumber \\
  && = \frac{1}{n} \cot \frac{\epsilon}{2n}
  \left\{
  - \int_{\epsilon}^{2\epsilon} d \theta~
  \ln \sin \frac{\theta}{2 n}
  + \int_{(2n-1)\pi}^{(2n-1)\pi+\epsilon} d \theta~
  \ln \sin \frac{\theta}{2 n}
  \right\}
\nonumber \\
  && = - \frac{1}{n} \cot \frac{\epsilon}{2n}
  \int_{\epsilon}^{2\epsilon} d \theta~
  \ln \sin \frac{\theta}{2 n}
  + 2 \ln \sin \frac{\pi}{2 n} + o(\epsilon).
\end{eqnarray}
The integral over $\widetilde{C}_n$ does vanish
in the limit $\epsilon \to 0$,
but in a slightly subtle way:
\begin{eqnarray}
  && \frac{2}{3}
  \int_{\pi+\epsilon/2}^{\pi-\epsilon/2} d \theta_3~
  \cot \frac{\epsilon}{3}~
  \ln \frac{\sin \frac{\theta_3+3\epsilon/2}{3}}
           {\sin \frac{\theta_3+\epsilon/2}{3}}
\nonumber \\
  && = \frac{2}{3} \cot \frac{\epsilon}{3}~
  \left\{
  \int_{\pi+2\epsilon}^{\pi+\epsilon} d \theta~
  \ln \sin \frac{\theta}{3}
  - \int_{\pi+\epsilon}^{\pi} d \theta~
  \ln \sin \frac{\theta}{3}
  \right\}
\nonumber \\
  && = 2 \left(
  \ln \sin \frac{\pi}{3}
  - \ln \sin \frac{\pi+\epsilon}{3} \right) + o(\epsilon)
  = o(\epsilon).
\end{eqnarray}
The final result for $V_{21/210+201}(n)$ is given by
\begin{eqnarray}
  V_{21/210+201}(n)
  &=& -\frac{(2n-1) \pi^2}{2 n^2}
  + 2 \left( \ln \frac{\sin \frac{\pi}{2 n}}
                      {\sin \frac{\epsilon}{2 n}} \right)^2
  + \frac{\pi^2}{6}
\nonumber \\
  && - \left( \ln \sin \frac{\epsilon}{2n} \right)^2
  - \left( \ln \sin \frac{\pi}{2n} \right)^2
  +2 \ln \sin \frac{\epsilon}{2n}~
  \ln \sin \frac{\pi}{2n} + I_n
\nonumber \\
  && - \frac{1}{n} \cot \frac{\epsilon}{2n}
  \int_{\epsilon}^{2\epsilon} d \theta~
  \ln \sin \frac{\theta}{2 n}
  + 2 \ln \sin \frac{\pi}{2 n}
  + o(\epsilon),
\end{eqnarray}
where
\begin{equation}
  I_{1} = \frac{\pi^2}{12}, \qquad
  I_{3/2} = \frac{4}{3}
  \int_{0}^{\pi} d \theta~
  \cot \frac{\theta+\pi}{3}
  \ln \sin \frac{\theta}{3}.
\label{I_n}
\end{equation}
\\

\noindent
{\bf Calculation of \boldmath $I_{3/2}$}

Let us calculate the integral $I_{3/2}$ in (\ref{I_n}).
It can be transformed to the following form
by change of variables:
\begin{eqnarray}
  I_{3/2}
  &=& -\frac{4}{3}
  \int_{-\pi/2}^{\pi/2} d \theta~
  \tan \frac{\theta}{3}~ \ln \sin
  \left( \frac{\theta}{3} + \frac{\pi}{6} \right)
\nonumber \\
  &=& -\frac{2}{3}
  \int_{-\pi/2}^{\pi/2} d \theta~
  \tan \frac{\theta}{3}
  \left[
  \ln \sin
  \left( \frac{\pi}{6} + \frac{\theta}{3} \right)
  - \ln \sin
  \left( \frac{\pi}{6} - \frac{\theta}{3} \right)
  \right]
\nonumber \\
  &=& -2 \int_{-1/\sqrt{3}}^{1/\sqrt{3}} dy~
  \frac{y}{1 + y^2} \ln \frac{1 + \sqrt{3} y}
                             {1 - \sqrt{3} y},
\end{eqnarray}
where $y$ in the last line is related to $\theta$
in the previous line by $y = \tan ( \theta / 3 )$.
The integral can be expressed in terms of polylogarithms.
We find
\begin{eqnarray}
  I_{3/2}
  &=& i \int_{-1/\sqrt{3}}^{1/\sqrt{3}} dy~
  \left( \frac{1}{1 - i y} - \frac{1}{1 + i y} \right)
  \ln \frac{1 + \sqrt{3} y}{1 - \sqrt{3} y}
\nonumber \\
  &=& 2 i \int_{-1/\sqrt{3}}^{1/\sqrt{3}} dy~
  \frac{1}{1 - i y}
  \ln \frac{1 + \sqrt{3} y}{1 - \sqrt{3} y}
\nonumber \\
  &=& -2 \int_{i - 1/\sqrt{3}}^{i + 1/\sqrt{3}} dy~
  \frac{1}{y} \left\{
  \ln \frac{1 - \sqrt{3} i}{1 + \sqrt{3} i}
  + \ln \left( 1 + \frac{\sqrt{3} y}{1 - \sqrt{3} i} \right)
  - \ln \left( 1 - \frac{\sqrt{3} y}{1 + \sqrt{3} i} \right)
  \right\}
\nonumber \\
  &=& \frac{4 \pi^2}{9}
  -2 \int_{1}^{1/2 - \sqrt{3} i/2} dy~
  \frac{1}{y} \ln (1-y)
  +2 \int_{1/2 + \sqrt{3} i/2}^{1} dy~
  \frac{1}{y} \ln (1-y)
\nonumber \\
  &=& \frac{4 \pi^2}{9} -4~ {\rm Li}_2 (1)
  +2~ {\rm Li}_2 ( e^{\pi i /3} ) +2~ {\rm Li}_2 ( e^{-\pi i /3} )
\nonumber \\
  &=& -\frac{2 \pi^2}{9}
  +2~ {\rm Li}_2 ( e^{\pi i /3} ) +2~ {\rm Li}_2 ( e^{-\pi i /3} ).
\end{eqnarray}
Using the formula\footnote{
This formula can be found, for example, at\\
$\qquad$
http://functions.wolfram.com/ZetaFunctionsandPolylogarithms/PolyLog2/03/01/
}
\begin{equation}
  {\rm Li}_2 ( e^{2 \pi i p / q} )
  = \frac{1}{q^2} \sum_{k=1}^{q-1} e^{2 \pi i k p / q}~
  \psi^{(1)} \left( \frac{k}{q} \right) + \frac{\pi^2}{6 q^2},
\end{equation}
where $p$ and $q ~ (0 < p \le q)$ are integers,
we can express
${\rm Li}_2 ( e^{\pi i /3} ) + {\rm Li}_2 ( e^{-\pi i /3} )$
as follows:
\begin{eqnarray}
  {\rm Li}_2 ( e^{\pi i /3} ) + {\rm Li}_2 ( e^{-\pi i /3} )
  = \frac{\pi^2}{108}
  -\frac{1}{18} \psi^{(1)} \left( \frac{1}{2} \right)
  +\frac{1}{36} \left\{ \psi^{(1)} \left( \frac{1}{6} \right)
                + \psi^{(1)} \left( \frac{5}{6} \right) \right\}
\nonumber \\
  -\frac{1}{36} \left\{ \psi^{(1)} \left( \frac{1}{3} \right)
                + \psi^{(1)} \left( \frac{2}{3} \right) \right\},
\end{eqnarray}
where $\psi^{(1)} (z)$ is defined by
\begin{equation}
  \psi^{(1)} (z) = \frac{d^2}{dz^2} \ln \Gamma (z).
\end{equation}
Since
\begin{equation}
  \psi^{(1)} (z) + \psi^{(1)} (1-z)
  = \frac{\pi^2}{\sin^2 \pi z},
\end{equation}
which follows from
\begin{equation}
  \Gamma (z) \Gamma (1-z) = \frac{\pi}{\sin \pi z},
\end{equation}
we have
\begin{equation}
  {\rm Li}_2 ( e^{\pi i /3} ) + {\rm Li}_2 ( e^{-\pi i /3} )
  = \frac{\pi^2}{18}.
\end{equation}
Therefore, the integral $I_{3/2}$ is given by
\begin{equation}
  I_{3/2} = -\frac{\pi^2}{9}.
\label{I_3/2}
\end{equation}
\\

\noindent
{\boldmath $V_{30} + V_{21} - V_{300} - V_{210} - V_{201}$}

Having finished all the preparations,
we obtain the following expression
for the sum of $V_{30/300}(n)$ and $V_{21/210+201}(n)$:
\begin{eqnarray}
  && V_{30/300}(n) + V_{21/210+201}(n)
\nonumber \\
  &=& \frac{2 \pi \ln 2}{\epsilon} - 6 \ln 2
  + \frac{\pi^2}{3} - (\ln 2)^2
  + \frac{(1-4n) \pi^2}{(2n)^2} + I_n
  + \left( \ln \frac{\sin \frac{\epsilon}{2n}}
                    {\sin \frac{\epsilon}{n}} \right)^2
  + o(\epsilon)
\nonumber \\
  &=& \frac{2 \pi \ln 2}{\epsilon} - 6 \ln 2
  + \frac{\pi^2}{3}
  + \frac{(1-4n) \pi^2}{(2n)^2} + I_n
  + o(\epsilon).
\end{eqnarray}
  From (\ref{V-to-V(n)}), (\ref{I_n}), and (\ref{I_3/2}),
$V_{30} + V_{21} - V_{300} - V_{210} - V_{201}$
is given by 
\begin{equation}
  V_{30} + V_{21} - V_{300} - V_{210} - V_{201}
  = -\frac{7 \pi^2}{36} + I_{1} - I_{3/2} + o(\epsilon)
  = o(\epsilon).
\end{equation}

\section{$V_{111}$}
\setcounter{equation}{0}

We calculate the integral
\begin{equation}
  \int_{\pi/2}^{3\pi/2} d \theta_3
  \int_{-\pi/2}^{\pi/2} d \theta_2
  \int_{-3\pi/2}^{-\pi/2} d \theta_1
  \left| 3 \sin \frac{\theta_2-\theta_1}{3} \right|^{-1}
  \left| 3 \sin \frac{\theta_3-\theta_1}{3} \right|^{-1}
  \left| 3 \sin \frac{\theta_3-\theta_2}{3} \right|^{-1},
\end{equation}
which is the value of $V_{111}$ (\ref{V_111})
in the limit $\epsilon \to 0$.
The integral
simplifies under the change of variables,
\begin{equation}
  t_i = \sqrt{3} \tan \frac{\theta_i}{3} \quad
  (i = 1, 2, 3),
\end{equation}
which corresponds to a conformal transformation
to the upper-half plane:
\begin{eqnarray}
  && \int_{\pi/2}^{3\pi/2} d \theta_3
  \int_{-\pi/2}^{\pi/2} d \theta_2
  \int_{-3\pi/2}^{-\pi/2} d \theta_1
  \left| 3 \sin \frac{\theta_2-\theta_1}{3} \right|^{-1}
  \left| 3 \sin \frac{\theta_3-\theta_1}{3} \right|^{-1}
  \left| 3 \sin \frac{\theta_3-\theta_2}{3} \right|^{-1}
\nonumber \\
  && = \int_{1}^{\infty} d t_3
  \int_{-1}^{1} d t_2
  \int_{-\infty}^{-1} d t_1~
  \frac{1}{(t_2-t_1)(t_3-t_2)(t_3-t_1)}.
\end{eqnarray}
The integrals with respect to $t_1$ and $t_2$
are easily carried out:
\begin{eqnarray}
  && \int_{1}^{\infty} d t_3
  \int_{-1}^{1} d t_2
  \int_{-\infty}^{-1} d t_1~
  \frac{1}{(t_2-t_1)(t_3-t_2)(t_3-t_1)}
\nonumber \\
  &=& \int_{1}^{\infty} d t_3
  \int_{-1}^{1} d t_2~
  \frac{1}{(t_3-t_2)^2} \ln \frac{t_3+1}{t_2+1}
\nonumber \\
  &=& \int_{1}^{\infty} d t_3~
  \left(
  \frac{1}{t_3-1} \ln \frac{t_3+1}{2}
  - \frac{1}{t_3+1} \ln \frac{t_3-1}{2}
  \right).
\end{eqnarray}
The integral over $t_3$ can be expressed in terms of
the polylogarithm ${\rm Li}_2 (z)$:\footnote{
The definition of the polylogarithm ${\rm Li}_2 (z)$
is given in (\ref{polylogarithm}).
}
\begin{eqnarray}
  && \int_{1}^{\infty} d t_3~
  \left(
  \frac{1}{t_3-1} \ln \frac{t_3+1}{2}
  - \frac{1}{t_3+1} \ln \frac{t_3-1}{2}
  \right)
\nonumber \\
  &=& \left[
  -2 ~{\rm Li}_2 \left( \frac{1-t_3}{2} \right)
  - \ln \frac{t_3-1}{2} \ln \frac{t_3+1}{2}
  \right]_{t_3=1}^{t_3=\infty}
\nonumber \\
  &=& \lim_{t_3 \to \infty} \left\{
  -2 ~{\rm Li}_2 \left( \frac{1-t_3}{2} \right)
  - \ln \frac{t_3-1}{2} \ln \frac{t_3+1}{2}
  \right\}.
\end{eqnarray}
Using the formula\footnote{
This formula can be found, for example, at\\
$\qquad$
http://functions.wolfram.com/ZetaFunctionsandPolylogarithms/PolyLog2/06/01/03/
}
\begin{equation}
  {\rm Li}_2 (-x) = -\frac{1}{2} \ln^2 (x) -\frac{\pi^2}{6}
                    + O \left( \frac{1}{x} \right)
  \quad {\rm for} \quad x > 1,
\end{equation}
we can calculate the limit:
\begin{equation}
  \lim_{t_3 \to \infty} \left\{
  -2 ~{\rm Li}_2 \left( \frac{1-t_3}{2} \right)
  - \ln \frac{t_3-1}{2} \ln \frac{t_3+1}{2}
  \right\} = \frac{\pi^2}{3}.
\end{equation}
Therefore, we have
\begin{equation}
  \int_{\pi/2}^{3\pi/2} d \theta_3
  \int_{-\pi/2}^{\pi/2} d \theta_2
  \int_{-3\pi/2}^{-\pi/2} d \theta_1
  \left| 3 \sin \frac{\theta_2-\theta_1}{3} \right|^{-1}
  \left| 3 \sin \frac{\theta_3-\theta_1}{3} \right|^{-1}
  \left| 3 \sin \frac{\theta_3-\theta_2}{3} \right|^{-1}
  = \frac{\pi^2}{3}.
\end{equation}


\renewcommand{\baselinestretch}{0.87}

\begingroup\raggedright\endgroup
\end{document}